\documentclass[fleqn,10pt]{wlscirep}
\usepackage[utf8]{inputenc}
\usepackage[T1]{fontenc}
\pdfoutput=1

\usepackage{titlesec}
\usepackage{parskip}
\usepackage{graphicx}	
\usepackage{amsmath}	
\usepackage{amssymb}	
\usepackage{multicol}        
\usepackage{bm}		
\usepackage{pdflscape}	
\usepackage{wasysym}
\usepackage{arydshln}
\usepackage{nameref}

\usepackage[resetlabels,labeled]{multibib}
\newcites{S}{Supplementary References}

\newcommand\0{\phantom{0}}
\newcommand{\cpd}{\,d$^{-1}$} 

\makeatletter
\def\ttl@useclass#1#2{%
  \@ifstar
    {\ttl@labeltrue\@dblarg{#1{#2}}}
    {\ttl@labeltrue\@dblarg{#1{#2}}}}
\makeatother

\title{Polarimetric detection of nonradial oscillation modes in 
\mbox{the $\beta\,$Cephei star $\beta$\,Crucis}}

\author[1,2,3,4,*]{Daniel V. Cotton}
\author[5,+]{Derek L. Buzasi}
\author[6,7,8]{Conny Aerts}
\author[3,9]{Jeremy Bailey}
\author[6]{\mbox{Siemen Burssens}}
\author[6,10]{May G. Pedersen}
\author[9]{Dennis Stello}
\author[4]{Lucyna Kedziora-Chudczer}
\author[3]{\mbox{Ain De Horta}}
\author[11]{Peter De Cat}
\author[9]{Fiona Lewis}
\author[9]{Sai Prathyusha Malla}
\author[4]{Duncan J. Wright}
\author[12]{Kimberly Bott}

\affil[1]{Monterey Institute for Research in Astronomy, 200 Eighth Street, Marina, CA, 93933, USA.}
\affil[2]{Anglo Australian Telescope, Australian National University, 418 Observatory Road, Coonabarabran, NSW 2357, Australia.}
\affil[3]{Western Sydney University, Locked Bag 1797, Penrith-South DC, NSW, 1797, Australia.}
\affil[4]{Centre for Astrophysics, University of Southern Queensland, Toowoomba, Queensland, 4350, Australia.}
\affil[5]{Department of Chemistry \& Physics, Florida Gulf Coast University, 10501 FGCU Boulevard S., Fort Myers, FL 33965, USA.}
\affil[6]{Institute of Astronomy, KU Leuven, B-3001 Leuven, Belgium.}
\affil[7]{Department of Astrophysics, IMAPP, Radboud University Nijmegen, 6500 GL Nijmegen, The Netherlands.}
\affil[8]{Max Planck Institute for Astronomy, K\"onigstuhl 17, 69117 Heidelberg, Germany.} 
\affil[9]{School of Physics, UNSW Sydney, New South Wales, 2052, Australia.}
\affil[10]{Kavli Institute for Theoretical Physics, Kohn Hall, University of California,
Santa Barbara, CA 93106, USA.}
\affil[11]{Royal Observatory of Belgium, Ringlaan 3, B-1180 Brussels, Belgium.}
\affil[12]{Department of Earth and Planetary Science, University of California, Riverside, CA, 92521, USA.}
\affil[ ]{}
\affil[*]{dc@mira.org, $^+$dbuzasi@fgcu.edu} 

\begin{document}

\flushbottom
\maketitle


\textbf{Here we report the detection of polarization variations due to nonradial modes in the $\beta\,$Cep star $\beta$\,Crucis. In so doing we confirm 40-year-old predictions of pulsation-induced polarization variability and its utility in asteroseismology for mode identification. In an approach suited to other $\beta\,$Cep stars, we combine polarimetry with space-based photometry and archival spectroscopy to identify the dominant nonradial mode in polarimetry, $f_2$, as $\ell=3$, $m=-3$ (in the $m$-convention of Dziembowski) and determine the stellar axis position angle as $25$ (or $205$)\,$\pm\,8^\circ$. The rotation axis inclination to the line of sight was derived as $\sim 46^\circ$ from combined polarimetry and spectroscopy, facilitating identification of additional modes and allowing for asteroseismic modelling. This reveals a star of $14.5\pm 0.5\,$M$_\odot$ and a convective core containing $\sim28\%$ of its mass -- making $\beta$\,Crucis the most massive star with an asteroseismic age.}


Asteroseismology has revolutionised our knowledge of low- and intermediate-mass stars across their entire evolution, determining fundamental parameters like mass, radius and age, and inferring interior rotation\cite{Hekker2017, Garcia2019, Aerts2019}. Some of the high-mass stars in the range 8--25\,M$_{\astrosun}$ are seismically active and known as $\beta$\,Cep stars\cite{Pamyatnykh1999}. Their pulsation modes are of low radial order and lack recognisable frequency patterns, making mode assignment difficult\cite{Aerts1998,Briquet2003}. Identification of the mode degree, $\ell$, and azimuthal order, $m$, is a prerequisite for asteroseismic inferences of the stellar interior \cite{Aerts2021}.

Because they are the progenitors of core-collapse supernovae and black holes, there is considerable interest in divining the interior structure of high-mass stars. However, identification of mode wavenumbers $(\ell,\,m)$ remains a large challenge for successful asteroseismology. Mode identifications in $\beta$\,Cep stars using multi-band photometry and light curve analysis \cite{Aerts2010} have proven difficult to achieve from ground-based data, with sometimes decades of effort expended on obtaining unambiguous mode identification in individual stars \cite{Aerts04}. Some progress on the asteroseismology of $\beta\,$Cep stars has resulted from extensive multisite campaigns \cite{Handler2006, Briquet2012} and from combined \textit{Microvariabilty and Oscillations of Stars} (\textit{MOST}) space photometry and ground-based spectroscopy \cite{Handler2009}. While such combinations of data were insightful for a few  $\beta\,$Cep stars, they remain too rare to advance the theory and understanding of their stellar interiors. As polarization is a pseudo-vector quantity, polarimetry potentially offers an advantage over other techniques, in that spatial information pertaining to the stellar surface is directly encoded in the signal through the polarization position angle (derived from normalised Stokes vectors $Q/I$ and $U/I$\cite{Stokes1851, Clarke10}).

Light scattering from electrons in the atmospheres of hot stars leads to linear polarization, being zero at the centre of the stellar disc but increasing radially to a maximum at the limb of the star, as first predicted by Chandrasekhar \cite{Chandrasekhar46}. The polarization will average to zero over a spherical star but can be observed if the star deviates from spherical symmetry. Distortion of a star due to nonradial pulsation\cite{Aerts2010} should therefore lead to temporal polarization variations. Models of these effects were first developed in the 1970s \cite{Odell79, Stamford80} and showed how polarization could provide constraints on the pulsation modes \cite{Watson83}. Most attempts to detect polarization signatures in $\beta$\,Cep stars have been unsuccessful \cite{Schafgans79, Clarke86, Elias08}, but Odell \cite{Odell81} reported a detection of polarization in the large amplitude $\beta$ Cep star BW\,Vul. However the result is questionable \cite{Aerts95} due to the small data set and the relative imprecision of the instrumentation of the time; it is also inconsistent with the identification of the pulsation as a radial mode \cite{Aerts95}, which would not produce any polarization.

Modern high-precision polarimeters \cite{Bailey15, Bailey20} can achieve much better precision than was possible in past work. These instruments have enabled the detection of long-predicted polarization effects caused by rapid rotation in hot stars \cite{Cotton17, Bailey20b} and shown B-stars to have a tendency for higher polarizations in general \cite{Cotton16}. In this project we used the HIgh Precision Polarimetric Instrument 2 (HIPPI-2) \cite{Bailey20} coordinated with photometric observations by the \textit{Transiting Exoplanet Survey Satellite} \textit{(TESS)} which is able to provide high-quality space photometry for pulsating B-stars \cite{Pedersen2019}. 

\section*{Results}
\label{sec:results}

Between 2018-Mar-29 and 2019-Jul-10 a total of 307 linear polarimetric observations were made of $\beta$\,Crucis ($\beta$\,Cru, Mimosa, HD\,111123), mostly at the 3.9-m Anglo Australian Telescope (AAT) at Siding Spring Observatory in the SDSS $g^{\prime}$ band (see Supplementary Figure \ref{fig:observations}). So configured HIPPI-2 can obtain better than 3~parts-per-million (ppm) precision in a short exposure (see \nameref{sec:methods} for details). The data include a number of sets of observations spanning half a night or more, these all show the polarization to vary smoothly in a periodic way with an amplitude of tens of ppm (see Figure \ref{fig:ts_examples}).

\textit{TESS} observed $\beta$\,Cru in 2-minute cadence for 27 days between 2019-Apr-22 to 2019-May-21 and thus overlaps the ground-based polarimetric observations in time (see \nameref{sec:methods}). Together these data sets, along with archival data from the \textit{Wide Field Infrared Explorer} (\textit{WIRE}), \textit{Hipparcos} and ground-based spectroscopy were used to carry out a joint frequency analysis (see \nameref{sec:methods}, Figure~\ref{fig:AS} and Supplementary Figure~\ref{fig:ASU}). Such an analysis identifies signals due to pulsations, but also other repeating signals that could be associated with rotation for instance.

The frequency analysis reveals (at $>4\sigma$ significance, i.e. a Signal-to-Noise Ratio, SNR, larger than 4) eleven distinct pulsation frequencies present in more than one data set, six more than previously identified for $\beta$\,Cru (a full list is given in Table \ref{tab:frequencies}). There are significant differences between the \textit{WIRE} and \textit{TESS} amplitudes, which may be indicative of time-varying mode amplitudes (see \nameref{sec:suppinf}) -- common for multiperiodic stars with heat-driven pulsations\cite{Pigulski2008}. These are the type of modes excited in the $\beta\,$Cep stars via the $\kappa$ mechanism \cite{Pamyatnykh1999}. Table \ref{tab:frequencies} also contains a column listing frequencies for fits to the combined \textit{WIRE}$+$\textit{TESS} data set. Any frequency appearing elsewhere in this is included in this column. Our motivation for including this column is that if frequencies and amplitudes are stable and not aliases, then the long combined time series has the potential to produce higher precision frequencies.

Amongst the detected frequencies are two in the polarimetric data corresponding to both photometric and spectroscopic frequencies. The first of these, which we detect in both linear Stokes parameters, is at 5.964\cpd and has been previously detected and assigned as $f_2$ \cite{Aerts1998, Buzasi2002, Cuypers2002}. In $Q/I$ and $U/I$ the amplitude of this mode is 7.08\,$\pm$\,0.91\,ppm and 5.78\,$\pm$\,0.97\,ppm respectively, roughly a hundredth of its $WIRE$ amplitude and a 35th of its \textit{TESS} amplitude.  All the detected modes of $\beta\,$Cru have lifetimes much longer than the individual data sets as revealed from Lorentzian fits to the power spectra and are thus confirmed to be undamped. The second frequency found in the polarimetric data is at 7.659\cpd ($f_6$) and is a newly found mode; we only detect it in $Q/I$ with an amplitude of 2.89\,$\pm$\,0.62\,ppm, about a 50th of its \textit{TESS} amplitude.

The uncertainty on the frequencies based on the polarimetry alone is an order of magnitude larger than the frequency determinations from \textit{TESS}, so we refit the polarimetric data as a sinusoid of the form $p=A \sin (2\pi f t + \phi)$ fixing the frequency to the \textit{TESS} value for $f_2 = 5.964(5)$\cpd. This produced amplitudes $A_Q=8.76\,\pm\,0.86$\,ppm and $A_U=7.97\,\pm\,0.94$\,ppm, and phase differences from the \textit{TESS} photometry of $\phi_Q-\phi=-0.817\,\pm\,0.113$\,rad and $\phi_U-\phi=-2.259\,\pm\,0.167$\,rad respectively, as shown in Figure \ref{fig:f2_phase}(a-c). Taking $2A$ as the polarimetric variation expected from $f_2$, this accounts for a third of the typical variability on a night (see Figure \ref{fig:ts_examples}), which suggests further modes may exist below our noise threshold. Alternatively, minor discrepancies in the calibrations between runs and/or inaccuracies in the zero point offsets as seen elsewhere\cite{bailey21}, could decrease the amplitudes.  

Pulsation-induced total light intensity, spectral line velocity, and polarization variations are driven by the cumulative effects of local changes in temperature, gravity and geometry due to the nonradial displacements caused by the modes. Watson \cite{Watson83}, whose notation we adopt in this paper, developed the most sophisticated analytical model of polarization in pulsating stars, following the same theoretical approach employed by Dziembowski \cite{Dziembowski77} for photometry.

We followed the procedures outlined by Watson \cite{Watson83} in applying the analytical model to $f_2$ -- the details are given in the \nameref{sec:methods}. In summary, we first noted that $\phi_U-\phi\simeq-\pi/2$, this is one of only two allowed values, indicating the polarization rotates anti-clockwise on the sky (N through E), and informing the sign of $m$. We next rotate the polarization between Stokes parameters so that in the model frame (denoted $M$) the phase difference $\phi_{Q\lambda}^M-\phi_\lambda$ is either $0$ or $\pi$. The angle of the rotation, $2\Phi_\chi$, is twice the position angle of the stellar axis on the sky (or twice the position angle plus $\pi$, since direction is ambiguous in polarimetry) -- the procedure is depicted in Figure \ref{fig:f2_phase}. Finally, the amplitudes of the fits to the pulsations in the rotated Stokes parameters, $A_{Q\lambda}^M$ and $A_{U\lambda}^M$, along with photometric amplitude, $A_{\lambda}^M$, eliminate some $(\ell,m)$ combinations and restrict the inclinations of others through the amplitude ratios $A_{U\lambda}^M/A_{Q\lambda}^M$ and $A_{Q\lambda}^M/A_{\lambda}^M$ as shown in Figure \ref{fig:f2_mode}.

As a consequence of symmetry, only $\ell\geq2$ modes can produce any significant polarization (see \nameref{sec:methods}), and modes with $\ell\geq5$ are not known to be prominent in photometry of slowly rotating $\beta$\,Cep stars \cite{Briquet2009, Handler2009b} -- and thus unlikely to be identified in the joint-frequency analysis; therefore we restricted our analysis to $2\leq\ell\leq4$ (see also \nameref{sec:suppinf} for further arguments). Then, as can be seen in Figure\,\ref{fig:f2_mode}, using the \textit{TESS} photometry, $f_2$ is limited to six possible modes, each associated with a distinct inclination range; these are: $(2,-2)$, $i=7 \substack{+4\\-2}^\circ$; $(3,+1)$, $i=66 \substack{+9\\-1}^\circ$; $(3,-2)$, $i=20 \substack{+8\\-4}^\circ$; $(3,-3)$, $i=37 \substack{+18\\-8}^\circ$; $(4,-2)$, $i=11 \substack{+5\\-3}^\circ$; $(4,-4)$, $i=27 \substack{+13\\-6}^\circ$. As described in the \nameref{sec:methods} we have cautiously used 3$\sigma$ errors to determine these ranges, and the position angles, which for the $(3,+1)$ solution corresponds to $\Phi_\chi=115\,\pm\,8^\circ$, and for the other modes to $\Phi_\chi=25\,\pm\,8^\circ$. This may still seem like a large number of possibilities, but the result allows us to very usefully refine our subsequent line profile analysis.

Stellar oscillations affect spectral line profiles dominantly through changes in the velocity field, although temperature changes may also be involved. The profile due to one mode, $p_{\rm spec}(\lambda,t)$, is described by four mode-dependent parameters, $\ell$, $m$, $A_\lambda$, $\phi_\lambda$, and two mode-independent parameters, $i$ and $v\sin i$\cite{Aerts2010}. Per additional mode and per wavelength bin $\lambda$, four extra free parameters occur in the expression for $p_{\rm spec}(\lambda,t)$ in the approximation of linear oscillation theory.  The inclination constraint from polarimetry offers a major reduction in the parameter space to consider for the spectroscopic mode identification. We carried out a line profile analysis based on the pixel-by-pixel method \cite{Zima2006} using FAMIAS \cite{Zima2008} to detect and identify the modes present in archival data of $\beta$\,Cru (see \nameref{sec:methods}) of the 4552.6\,\AA\:Si\,III line. Eight frequencies from Table \ref{tab:frequencies} have significant amplitudes and phases across the line profiles -- these are shown in Figure\,\ref{fig:lpa}.

Three modes have amplitudes dominant over the other five in the spectroscopic data; they are frequencies $f_2$, $f_8$, and $f_1$ which are detected at $>10\sigma$. Nevertheless, we performed mode identification for all eight modes, initially adopting the restriction $i<75^\circ$. We then eliminated those solutions that were inconsistent with the polarimetric results; first by restricting the inclination to the ranges allowed for $f_2$ as described above (Figure\,\ref{fig:inclvsini} shows how applying this constraint alone cuts down the allowed solutions significantly), then subsequently to accommodate the detection of $A^M_U$ in $f_6$ and the non-detections of other frequencies in polarimetry. The outcome of the combined mode identification is demonstrated in Supplementary Tables\,\ref{tab:MI-all8} and \ref{tab:MI-best} and reveals a fully consistent solution for $(\ell_2,m_2)=(3,-3)$ and $v\sin i=14\pm 2$\,km\,s$^{-1}$. The strongest mode in photometry, $f_1$, is found to be a dipole mode in agreement with previous results\cite{Briquet2003}. 
We were also able to place restrictions on the mode identification of $f_8$ and the five lower-amplitude modes detected in the spectroscopy, although the uniqueness for those is not guaranteed as several solutions are almost equivalent (see \nameref{sec:methods}). For the mode with frequency $f_6$, a combination of spectroscopy and polarimetry delivered additional information (see the \nameref{sec:suppinf}, including Supplementary Figure\,\ref{fig:f6_phase}, for details of its assignment and also constraints placed on the remaining frequencies by their non-detection in polarimetry).

Few $\beta\,$Cep stars have been modelled asteroseismically with inferred values of the mass, radius, age, and convective core mass, due to lack of mode identification. We performed such initial modelling for $\beta$\,Cru, based on a neural network trained for radial, dipole, and quadrupole modes\cite{HendriksAerts2019}. We applied this network to the star's two lowest degree $\ell\leq 2$ modes ($f_1$ and $f_5$). The results are presented in Supplementary Figures\,\ref{fig:HRD} and \ref{fig:NN} and reveal a 11.3$\pm$1.6 million year old star of $14.5\pm0.5\,$M$_\odot$ with a central hydrogen mass fraction in the range $[0.30,0.33]$, having a convective core of $\sim 28\pm3\%$ of its total mass and slow surface rotation with a period between 13 and 17\,d. The rotation period is not detected in the space photometry, possibly because it falls in the low-frequency regime of the spectrum (see Figure\,\ref{fig:AS}).

\section*{Discussion}
\label{sec:discussion}

Previously, it has proved difficult to measure stellar pulsations polarimetrically. Here, using the most precise optical astronomical polarimeter available, we have been able to detect the signatures of two modes in $\beta$\,Cru with small polarimetric amplitudes. A number of further modes are predicted to have polarimetric amplitudes just below the detection threshold, which is motivation for extending our observations and also for developing more sensitive instruments. Additionally, the calculations presented in Figure \ref{fig:f2_mode} demonstrate that in some cases, favourable inclination and mode combinations could produce very large polarisation amplitudes indeed -- this is particularly true for higher-order modes. Thus, even polarimetry at a precision of 10s of ppm should be sufficient to restrict the allowed modes and/or inclination in other similar stars.

With the polarimetric data presented in the paper, $\beta\,$Crucis is currently the highest-mass $\beta\,$Cep star with such asteroseismic information.  While these results can be improved by dedicated modelling including the interior rotation from the identified modes, our study reveals that combined time-series space photometry along with ground-based spectroscopy and polarimetry leads to a self-consistent solution for the modes and their identification. A crucial aspect for future applications to stars without line-profile variability data is that the polarimetry places stringent constraints on the inclination angle of the star. This is demonstrated by Figure \ref{fig:inclvsini}, which shows the reduction in the number of allowed solutions associated with just the polarimetric constraints from one mode, $f_2$.

The relative phases of the modes in polarimetry and photometry were crucial for applying the analytical model to $f_2$. Of the space-based photometric missions that have observed $\beta$\,Cru, only the \textit{TESS} observations were sufficiently contemporaneous to facilitate this determination. \textit{TESS}'s improved precision over past photometric missions produced additional frequency identifications, and as per our mode identification, many of these frequencies are likely to be higher order modes, to which polarimetry is more sensitive generally. Therefore, if polarimetry is to be used to maximal effect in mode determination, it is urgent to acquire similar polarimetric data on other $\beta$\,Cep stars being observed by \textit{TESS} during its mission, which has been extended through 2022\cite{coffey2020tess}.

There are a dozen known $\beta$\,Cep stars brighter than $m_V=5$ with pulsation amplitudes larger than or similar to $\beta$\,Cru\cite{Stankov05} to which this method can now be applied. Once a sufficient number of their interior structures are determined, main-sequence stellar evolution calculations can be calibrated and more accurately extrapolated to the stage prior to the core collapse, which in turn informs the spectral and chemical evolution theories of galaxies\cite{Bernardi2003}.


\section*{Methods}
\label{sec:methods}
\phantomsection
\subsection*{Archival time-series photometry and spectroscopy}

Extensive time-series studies of $\beta\,$Cru have been performed based on high-resolution ground-based spectroscopy spread over 13 years \cite{Aerts1998} and on \textit{WIRE}\footnote{\url{https://www.jpl.nasa.gov/missions/wide-field-infrared-explorer-wire/}} space photometry \cite{Buzasi2002, Cuypers2002}. \textit{WIRE} observed $\beta$\,Cru in June - July 1999 for a total of 17 days with a cadence of 10~Hz. For the current re-analyses, those \textit{WIRE} data were re-reduced using the \textit{WIRE} pipeline developed later in the mission\cite{Bruntt2005} and binned to 120~s cadence. We refer to\cite{Aerts1998} for details on the 13-yr long spectroscopic data set of high-resolution spectroscopy, consisting of 1193 high-resolution high-SNR spectra. These data led to the discovery of the spectroscopic binary nature of $\beta\,$Cru, with an orbital period of 5\,yr and a B2\,V secondary. Clear line-profile variations were discovered and led to the detection of three oscillation frequencies. These are included in Table\,\ref{tab:frequencies} and discussed further below. The combined \textit{WIRE} and spectroscopic high-precision data analyses led to the conclusion \cite{Cuypers2002} that the two dominant modes in the \textit{WIRE} photometry with frequencies $f_1\simeq 5.231$\cpd and $f_3\simeq 5.477$\cpd are different from the dominant mode in the second moment of the line-profile variability, $f_2\simeq 5.959$\cpd. These three modes were found to be nonradial $(l\neq 0)$, but aside from $\ell_1=1$ the identifications of $\ell_2$ and $\ell_3$, as well as all of $m_1, m_2, m_3$ remained ambiguous.

\subsection*{Polarimetric observations}

We have made 307 high precision polarization observations of $\beta$\,Cru using the 3.9-m Anglo Australian Telescope (AAT) at Siding Spring and the 60-cm telescope at Western Sydney University's (WSU) Penrith Observatory using the HIgh Precision Polarimetric Instrument 2 (HIPPI-2 \cite{Bailey20}); these observations were made between March 2018 and July 2019. A single observation was also made at UNSW Observatory in Sydney with a 35-cm Celestron C14 telescope using Mini-HIPPI \cite{Bailey17} in May 2016. 

All the observations used Hamamatsu H10720-210 photo-multiplier tubes as the detectors. The operating procedures were as standard \cite{Bailey20}, the instrument was always mounted at the Cassegrain focus, and the instrument/telescope configurations details are summarised in Supplementary Table \ref{tab:observations}. A single sky (S) measurement was made adjacent to each target (T) measurement at each of the four position angles, $PA=0, 45, 90, 135^{\circ}$, in the pattern TSSTTSST. On a number of occasions the observations were made back-to-back in a series stretching a few to several hours.

A small polarization due to the telescope mirrors, TP, shifts the zero-point offset of our observations. This is corrected for by reference to the straight mean of several observations of low polarization standard stars, details of which are given in Supplementary Table \ref{tab:tp}. The position angle is calibrated by reference to literature measurements of high polarization standards \cite{Bailey20} listed in Supplementary Table \ref{tab:pa}. 

Most of the observations were made using an SDSS $g^{\prime}$ filter in which HIPPI-2 has a nominal (ultimate) precision per observation, $e_P$ of 1.7 ppm on the AAT and 8.2 ppm on the WSU 60-cm telescope. Without a filter (Clear) on the AAT this figure is 2.9 ppm; the equivalent figure for Mini-HIPPI on the UNSW 35-cm is 14.0 ppm\cite{Bailey20}. To determine the error for any given measurement we take the square root of the sum of $e_P$ and the internal measurement error, $\sigma_P$, which is the standard deviation of the polarization determined from each integration that makes up a measurement. Exposure times are selected so that $\sigma_P$ is similar to $e_P$. Some observations were made in poor weather, which has the effect of raising $\sigma_P$ -- from a lower photon count -- and/or increasing the dwell time per observation (when scattered clouds are allowed to pass before beginning a new measurement). The dwell time is the total time for an observation, including sky exposures and dead time, between the start of the first measurement and the end of the last measurement in the PA sequence.

Without a filter, the instrument has a similar effective wavelength ($\lambda_{\rm eff}$) to the SDSS $g^{\prime}$ filter, but a much wider band: $\sim$350~to~700~nm, compared to $\sim$400~to~550~nm. However, both the intrinsic and interstellar polarization will have a wavelength dependence, so the subsequent frequency and mode analysis is limited to the 274 SDSS $g^{\prime}$ observations. The small variability in $\lambda_{\rm eff}$, and the modulation efficiency (Eff) within a run (as shown in Supplementary Table \ref{tab:observations}) is purely a result of taking account of airmass differences in the bandpass model. All of the observations were reduced using the standard procedures for HIPPI-class instruments \cite{Bailey20}.

\subsection*{New \textit{TESS} photometry}

NASA's \textit{Transiting Exoplanet Survey Satellite} (\textit{TESS}) Mission \cite{Ricker2014} began performing a 2-year near-all-sky survey in 2018 as part of its prime mission. The survey covered each hemisphere in 13 sectors, each lasting roughly 27 days, though there is significant sector overlap near the ecliptic pole so that some portions of the sky had longer periods of near-continuous coverage. Each sector was observed with 30-minute cadence, but a subset of pre-selected targets were observed at a shorter 2-minute cadence. 

\textit{TESS} observed $\beta$~Cru in 2-minute cadence for 27 days during Sector 11 of Cycle 1; the observing dates for that sector cover 2019 Apr 22 to 2019 May 21 and thus overlap the ground-based polarimetric observations in time. We produced a light curve using the target pixel files produced by the \textit{TESS} Science Processing Operations Center (SPOC \cite{Jenkins2017}) using the procedure outlined in Nielsen et al \cite{Nielsen2020}. Effectively we produced time series for each pixel and then derived an aperture photometry mask which minimized the mean absolute deviation figure of merit ${q = \sum_{i=1}^{N-1}\mid f_{i+1}-f_{i}\mid}$, where $f_i$ is the flux at cadence $i$, and $N$ is the length of the time series. The resulting light curve was then detrended against the centroid pixel coordinates by fitting a second-order polynomial with cross terms, resulting in a light curve with modestly improved noise characteristics compared to the SPOC product. We note that, while the \textit{TESS} image is saturated, the large postage stamp of 549 pixels effectively captures all electrons which overflow down the columns, and the stamp is far enough (200 pixels) from the edge of the detector to ensure that no electrons are lost. The \textit{TESS} data release notes\footnote{\url{https://archive.stsci.edu/missions/tess/doc/tess_drn/tess_sector_11_drn16_v02.pdf}} do not identify this target as problematic, and we have confirmed that the count rate is within 1\% of the level expected for a \textit{TESS} magnitude of $2.82$.

\subsection*{Joint frequency analysis}

Our frequency analysis was motivated by the desire to take advantage of the numerous mutually supporting data sets available, rather than to try to analyze each fully independently and later combine those results. This joint analysis made use of light curves from \textit{WIRE}, \textit{TESS}, $g^{\prime}$ band polarimetry, archival ground-based spectroscopy, and archival \textit{Hipparcos} data. In each case we corrected times to BJD-2440000 and applied a high-pass filter to truncate frequencies below $\sim 1\,\rm d^{-1}$. Since we wished to conduct a joint analysis, and the different observation sets measured different physical quantities (relative flux in three different bands, polarization, and velocity moments $\langle v\rangle$ and $\langle v^2\rangle$), we scaled each light curve by dividing by its standard deviation to ensure that the relative dispersions of each were similar. 

The algorithm we applied began with calculation of discrete Fourier transforms (DFT) and amplitude spectra for each scaled light curve. In each case we used the same frequency grid $[0, 10]\,\rm d^{-1}$ with frequency resolution sufficient to $10\times$ oversample the longest time series. We then followed the methodology of Sturrock et al.\cite{Sturrock2005} in constructing a joint amplitude spectrum from the two highest SNR photometric time series, \textit{WIRE} and \textit{TESS}, as
\begin{equation}
X =  \left(\frac{2}{N_W} \times \frac{2}{N_T} \right)^{1/2} \left(F_W \circ F_T \right)^{1/2}
\end{equation}
where $N_W$ and $N_T$ represent the number of points in the \textit{WIRE} and \textit{TESS} time series, respectively, and $F_W$ and $F_T$ the corresponding Fourier transforms. An advantage to this approach is its ability to handle alias and instrumental peaks; as these are different in each time series, their amplitudes are depressed in the joint spectrum.

The largest amplitude peak was identified in the joint amplitude spectrum and used as a starting point for sine curve fits to each individual time series, including a fit to the full combined \textit{WIRE}+\textit{TESS} light curve. Each time series was then prewhitened by the frequency determined from its specific sinusoidal fit, a new joint \textit{WIRE}+\textit{TESS} amplitude spectrum was constructed, and the process repeated until 50 frequencies had been identified and removed. 

For this study, we then conservatively retained for further analysis only those frequencies for which SNR $>4$ for two or more individual time series, as shown in Table \ref{tab:frequencies}. This process is intentionally conservative, with the goal of building a minimal rather than a maximal frequency list. Thus, the $6.877\,\rm d^{-1}$ peak visible in Figure \ref{fig:AS}(b) does not appear in Table \ref{tab:frequencies} because it is only significant in the \textit{TESS} data set (the frequency is detected in the \textit{WIRE} light curve, but only with SNR $\sim 2.5$). Similarly, for the \textit{Hipparcos} time series, while we do detect $f_3$, whose presence was reported as ``marginal'' by Cuypers et al.\cite{Cuypers2002}, it does not appear in our table because it fails to meet our SNR criterion.

Each individual light curve also produces a number of frequencies that fail to satisfy our SNR requirement. The majority of these are low-frequency peaks ($< 1\,\rm d^{-1}$) which are presumably either due to a stellar process such as internal gravity waves, rotation, or near-core convection, or to instrumental effects, and which are unique to a particular time series. Thus, in order to identify all of the peaks in each individual data set which satisfy our SNR requirement, we remove a total of 23 frequencies. The residual amplitude spectrum shown in Figure \ref{fig:AS}(c) is the result of removing those 23 frequencies from the TESS light curve. 

We also confirmed by inspection of the final prewhitened amplitude spectra for the individual time series that no meaningful peaks remained in any of these. Finally, we fit in a least-squares sense a multicomponent sinusoidal model to each data set using the frequencies, amplitudes, and phases from the above algorithm as input values which are allowed to vary within a narrow range, and the results of that fit are used to finalize frequency determinations and estimate uncertainties. Using this process, we achieve improved fits to the first and second moment variations\cite{Aerts1998} thanks to the frequencies established from the combined space photometry.

\subsection{Polarimetric mode identification}

To make mode assignments using polarimetry we employ the analytical model of Watson \cite{Watson83}. This model assumes that deviations from spherical symmetry are small; it takes account of the effect of local changes in temperature and gravity on the light intensity, the effect of surface twist on changing $Q$ and $U$ in the plane of the sky, and the changing projected area of a surface element. It neglects rotation effects, the effect of changes in the local temperature and gravity on the limb darkening function and changing polarization resulting from the changing surface normal.

In the reference frame of the stellar axis projected on the plane of the sky, the model\cite{Watson83} (denoted by $^M$) uses the ratio of polarimetric amplitudes ($A_{U\lambda}^M/A_{Q\lambda}^M$) and the ratio of polarimetric to photometric amplitudes ($A_{Q\lambda}^M/A_\lambda$), along with the relative phases ($\phi_{U\lambda}^M-\phi_{Q\lambda}^M$ and $\phi_{Q\lambda}^M-\phi_\lambda$) to constrain the allowed modes (or uniquely identify a mode if the inclination, $i$, is precisely known). The wavelength, $\lambda$, dependent amplitudes depend primarily on geometry (although the polarimetric amplitudes are wavelength dependent, their ratio is not); the detail of the atmosphere is accounted for through only two parameters, the polarimetric scaling factor $z_{\ell\lambda}$ and photometric scaling factor $b_{\ell\lambda}$, determined by radiative transfer calculations (see the next \nameref{sec:methods} section). Note that the wavelength dependence of the phases comes about only as a result of sign changes in the ratio $z_{\ell\lambda}/b_{\ell\lambda}$. 

\begin{figure*}
\begin{center}
\includegraphics[width=\textwidth, bb = 0 0 9in 8.15in]{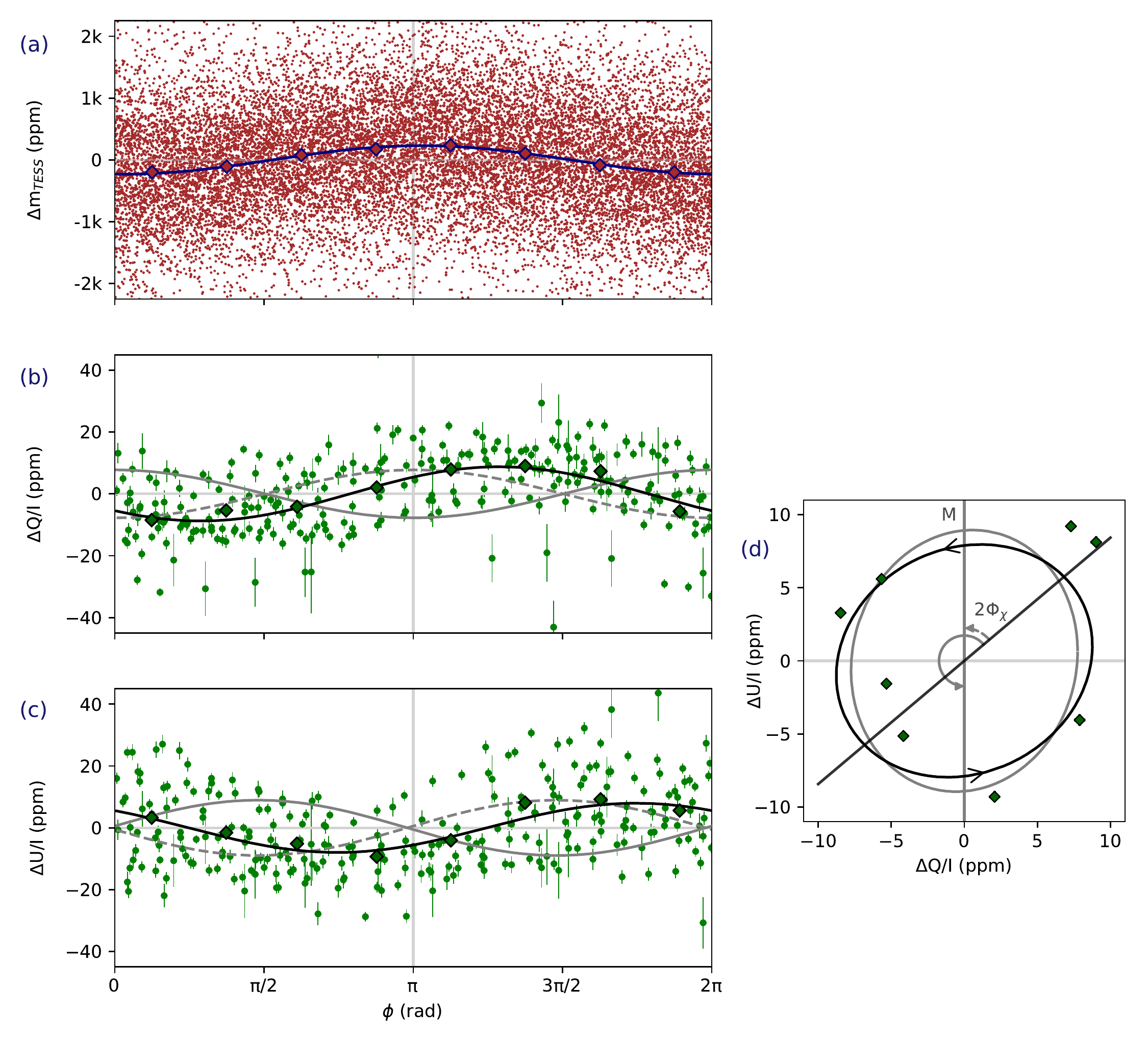}
\end{center}
\caption{\textbf{Photometric and polarimetric data phase-folded to $f_2$ for $\beta$\,Cru |} In panels (b) and (c) the data are shown as the change in $Q/I$ and $U/I$ from the error weighted means (266.5 $\pm$ 0.2 ppm and 3.1 $\pm$ 0.2 ppm) as green dots (the plotted uncertainties are the nominal 1-$\sigma$ errors, which are a combination of photon shot noise and an instrumental `positioning error' \cite{Bailey20}). Similarly, photometric data from \textit{TESS} are shown as brown dots in panel (a). The sinusoidal fits to the polarimetric and photometric data, locked at the \textit{TESS}-derived frequency $f_2 = 5.964$\,d$^{-1}$, are shown as the solid black and blue lines, respectively. The diamonds in each plot represent the data binned in increments of $\pi/4$\,rad. Panel (d) shows the phase-binned polarimetric data as a Q-U diagram (the formal errors are the size of the data points). Projected onto this plane the sinusoidal fit becomes a polarization ellipse; the black arrows indicating the direction of data progress with phase, which is the same as it is on the sky. Rotating the ellipse so that major and minor axis correspond to the $\Delta U/I$ and $\Delta Q/I$ axes (or, in general, vice versa) aligns the data to the frame of the analytical model of Watson \cite{Watson83}, labelled $^M$, and shown in grey. The rotation angle required is twice the position angle of the stellar axis ($\Phi_\chi$). The two allowed solutions correspond to $\phi_{Q\lambda}^M-\phi_\lambda=0$ (dashed grey line) and $\phi_{Q\lambda}^M-\phi_\lambda=\pi$ (solid grey line) in panels (b) and (c). The values of $A^M_{Q\lambda}$ and $A^M_{U\lambda}$ then correspond to the intersection of the rotated polarization ellipse with the axes of the Q-U diagram.}
\label{fig:f2_phase}
\end{figure*}

To determine these quantities for $f_2$ we rotate the measured polarization to align with the model frame, as demonstrated in Figure\,\ref{fig:f2_phase}. Algebraically, such rotation is accomplished, in general, by $Q^\prime=Q\cos{2\theta}-U\sin{2\theta}$ and $U^\prime=U\cos{2\theta}+Q\sin{2\theta}$, where $2\theta$ is the rotation angle. In this instance the rotation must result in $\phi_{Q\lambda}^M-\phi_\lambda=0$ or $\pi$, with each solution corresponding to different modes. This results in $\theta$ corresponding to the stellar axis polarization position angle, $\Phi_\chi$ -- measured North over East, in this case $25$ or $115 \pm 3^\circ$ (1$\sigma$ errors). Because polarization is a pseudo-vector -- it has a magnitude and \textit{orientation} as opposed to a vector, which has a magnitude and a direction -- the true position angle of the stellar axis will be either $\Phi_\chi$ or $\Phi_\chi+\pi$, meaning it may \textit{also} be $205$ or $295 \pm 3^\circ$. In the figure $\phi_U-\phi_Q\approx-\pi/2$  ($-1.44 \pm 0.24$\,rad) -- only a value of $\pi/2$ or $-\pi/2$ is allowed -- and this value means the polarization rotates counter-clockwise (N through E) on the plane of the sky with time, which determines the sign of $m$ for a given $\ell$ and $i$. $A_{Q\lambda}^M$ and $A_{U\lambda}^M$ correspond to the amplitudes of the rotated Stokes parameters in Figure\,\ref{fig:f2_phase}(b) and (c).

In the case of a radial mode, the mode symmetry results in no net polarization change over the stellar disc. Less obviously, under the assumptions of the analytical model, a dipole mode ($\ell=1$) also produces no net polarization; this is a consequence of polarization measuring orientation rather than direction, and thus distortions at opposite hemispheres cancelling \cite{Odell79, Stamford80, Watson83}. Only modes of order $\ell=2$ and $3$ have previously been calculated using the analytical model. Following Watson's \cite{Watson83} approach we also made calculations for $\ell=4$, which we present here; the work required additional angular momentum transformation matrix elements, which are given in Supplementary Table \ref{tab:matrix-elements}. Despite potentially larger polarimetric amplitudes, larger $\ell$ values are less likely to be detected, as a consequence of smaller photometric amplitudes (see \nameref{sec:suppinf}). Therefore we infer $f_2$, and indeed any mode, detected in $\beta$\,Cru here with polarimetry must have $\ell = 2, 3$ or $4$. Using the measured amplitudes plotted against the model predictions in Figure\,\ref{fig:f2_mode} we can place tighter constraints on the allowed modes in terms of $\ell$, $m$ and $i$. The results are shown for both the \textit{TESS} ($\lambda$ = 800~nm) and \textit{WIRE} ($\lambda$ = 600~nm) photometric amplitudes. Given the decade-long gap between the polarimetric and \textit{WIRE} data sets, we rely on the \textit{TESS} phasing for both. Using the \textit{TESS} photometry the allowed modes and inclinations (3$\sigma$ errors) are: $(2,-2)$, $i=7 \substack{+4\\-2}^\circ$; $(3,+1)$, $i=66 \substack{+9\\-1}^\circ$; $(3,-2)$, $i=20 \substack{+8\\-4}^\circ$; $(3,-3)$, $i=37 \substack{+18\\-8}^\circ$; $(4,-2)$, $i=11 \substack{+5\\-3}^\circ$; $(4,-4)$, $i=27 \substack{+13\\-6}^\circ$ . The $(3,+1)$ solution corresponds to $\Phi_\chi=115 \pm 8^\circ$, the others to $\Phi_\chi=25 \pm 8^\circ$. We conservatively adopted 3$\sigma$ errors here for $i$ and $\Phi_\chi$ to allow for $\phi_U-\phi_Q$ not being exactly $-\pi/2$, the possibility that the amplitudes are damped by zero-point or inter-run calibration offsets, as well as effects not included in the analytical model, and the fact that our calculations are monochromatic but the observations broadband.

\subsection*{Polarized radiative transfer}

The analysis of Watson \cite{Watson83} provides expressions for the amplitude of pulsations in intensity and polarization that depend on the details of the stellar atmosphere model only through a pair of values (for a given $\ell$ and wavelength $\lambda$) $b_{\ell\lambda}$ and $z_{\ell\lambda}$ which are derived from the viewing angle ($\mu = \cos{\theta}$) dependence of the modelled intensity and polarization. A small number of these values, mostly for ultraviolet wavelengths are tabulated by Watson \cite{Watson83}. We calculated new values using {\tt ATLAS9} stellar atmosphere models and the {\tt SYNSPEC/VLIDORT} stellar polarization code \cite{Cotton17, Bailey19, Bailey20b} which is a version of the {\tt SYNSPEC} spectral synthesis code \cite{Hubeny85} that we have modified to do polarized radiative transfer using the {\tt VLIDORT} code\cite{Spurr06}. As a check we recalculated the values listed by Watson \cite{Watson83} for a star of $T_{\rm eff}$ = 23000\,K and $\log{g}$ = 3.6 and found good agreement. We calculated new values for a star of $T_{\rm eff}$ = 27000\,K and $\log{g}$ = 3.6 appropriate for $\beta$\,Cru \cite{Morel2008}. The calculated values are given in Supplementary Table \ref{tab:zl_bl}.

\subsection*{Mode identification from combined spectroscopy and polarimetry}

We revisited the archival time-series spectroscopy, described in detail by Aerts et al.\cite{Aerts1998}; consisting of 1193 high-resolution (binned to $\Delta\lambda=0.03$\AA) high signal-to-noise (SNR$\in [250,750]$) spectra covering the Si\,III$\lambda\lambda\,4552,4568.4574$\AA\ triplet. The time base spans 13 years but the temporal coverage is sparse; the data was taken on 25 individual nights, with intense monitoring during a single night twice, and on two, three, and four consecutive nights once each. These data led to the discovery of the spectroscopic binarity of the star, revealing an eccentric system with an orbital period of $\sim\!5$\,yr and a B2\,V secondary with $T_{\rm eff}=22\,000$K, $\log\,g\simeq 4.0$, delivering $\sim\! 8\%$ of the flux\cite{Aerts1998} (see the \nameref{sec:suppinf} for additional information on the companion). 

Here, we used the orbit-subtracted line profiles of the Si\,III triplet to perform mode identification with the pixel-by-pixel method, also known as the Fourier Parameter Fit method \cite{Zima2006}. This method had not yet been applied to $\beta\,$Cru's complex line-profile variability. It can handle more modes than the moment method, because it is based on the periodic variability in each of the wavelength bins belonging to the spectral lines, rather than using integrated quantities across the profiles as is the case for the moment method\cite{Telting1997}.  We fixed the frequencies found in the space photometry according to Table\,\ref{tab:frequencies} to test which of those have significant amplitude $A_\lambda$ and phase $\phi_\lambda$ across each of the three Si\,III lines. The outcome is consistent for each of the three lines of the Si triplet, and are displayed in Figure\,\ref{fig:lpa} for the deepest of the three lines. Its central laboratory wavelength occurs at 4552.6\AA\ and the variability in this line reveals eight of the eleven frequencies as significant, labelled as such in Table\,\ref{tab:frequencies}. 

Spectroscopic mode identification with the Fourier Parameter Fit method using the code {\tt FAMIAS}\cite{Zima2008} was performed for the eight modes by treating them separately from their amplitude and phase behaviour shown in  Figure\,\ref{fig:lpa}, adopting a grid-based approach. {\tt FAMIAS} relies on a user-provided value of the effective temperature and gravity of the star to predict its limb darkening for the considered spectral line.
For each of the eight modes, we varied the degree in $\ell\in [0,4]$ and the azimuthal order $m\in [-\ell,\ell]$. This mode identification method is known to provide more robust results for the azimuthal order than for the mode degree, particularly for zonal modes revealing a step-like behaviour in the phase across the line profile\cite{Aerts2010}.
The local amplitudes of the modes were allowed to vary from 1 to 50\,km\,s$^{-1}$ in steps of 1\,km\,s$^{-1}$, while the inclination angle of the rotation axis varied from $2^\circ$ to $75^\circ$ in steps of $2^\circ$ and the projected rotation velocity $v\sin i$ from 10 to 40\,km\,s$^{-1}$ in steps of 2\,km\,s$^{-1}$. An additional free parameter needed to fit line profiles is the so-called intrinsic line broadening, here approximated as a Gaussian with width ranging from 10 to 30\,km\,s$^{-1}$ in steps of 2\,km\,s$^{-1}$. We then find the acceptable solutions for the spectroscopic mode identification by extracting those with one common value for $i$ and $v\sin i$. This produces thousands of options, some of which are listed in Supplementary Table\,\ref{tab:MI-all8}. The results for $i$ and $v\sin i$ for all the solutions having $\chi^2\leq 1000$ are shown in Figure\,\ref{fig:inclvsini}. Only solutions with $\ell=3$ or $4$ for the dominant mode in polarimetry remain, in full agreement with spectroscopic mode identification from the moment method\cite{Briquet2003}.
 
\begin{figure*}
\begin{center}
\includegraphics[clip, trim={0cm, 5.5cm, 0cm, 5.cm}, width=\textwidth]{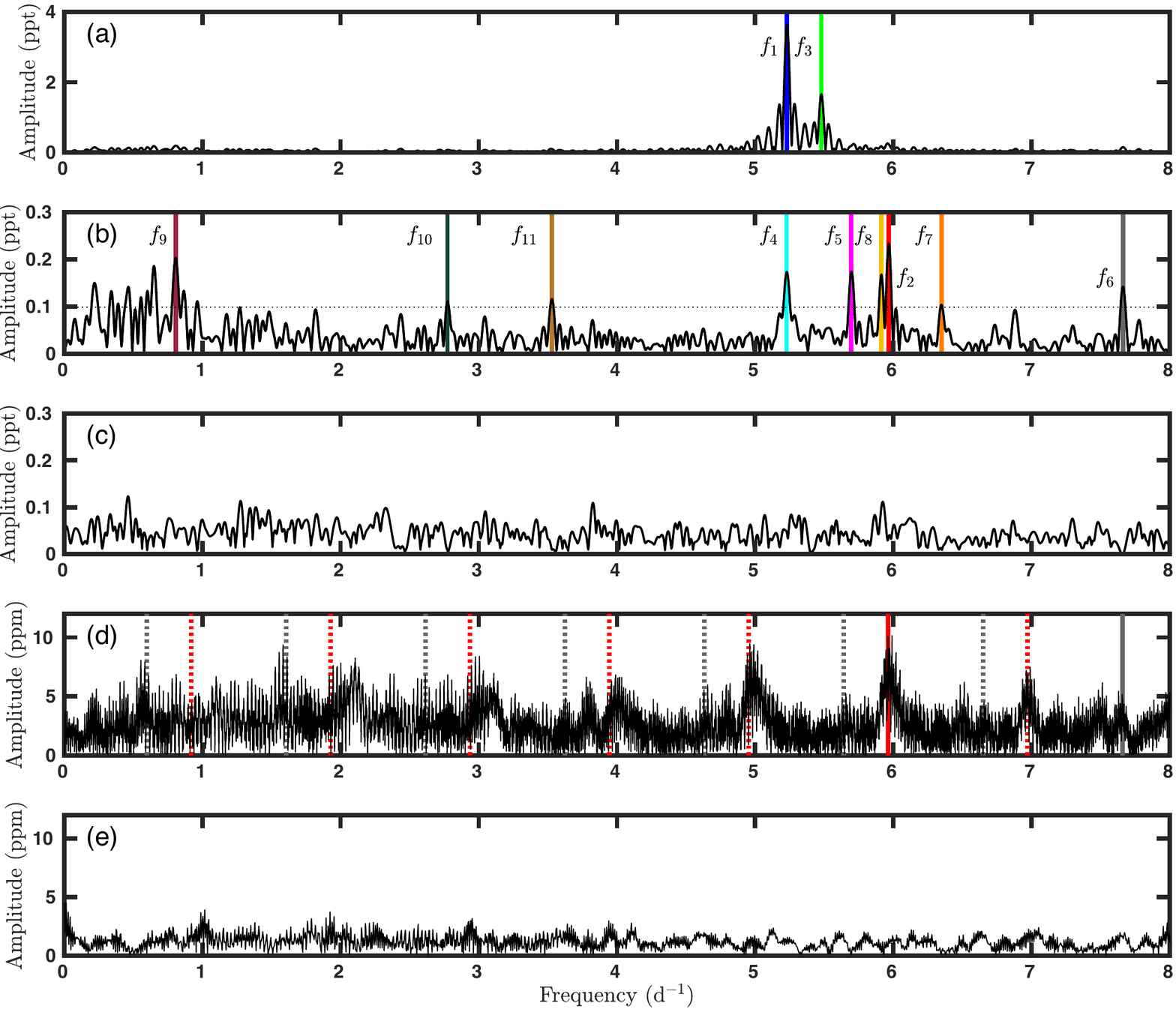}
\end{center}
\caption{\textbf{Selected amplitude spectra for $\beta\,$ Cru |} 
Panel (a) shows the amplitude spectrum derived from the \textit{TESS} time series for $f \in [0,8] \rm~d^{-1}$. Coloured vertical lines illustrate the two oscillation modes with the largest amplitudes; other frequencies are not obvious in the amplitude spectrum until prewhitening has been performed. Panel (b) shows the amplitude spectra derived from the \textit{TESS} time series in black, after prewhitening for $f_1$ and $f_3$. Peak locations are identified by coloured vertical lines. Panel (c) shows the \textit{TESS} amplitude spectrum after 23 identified modes have been prewhitened, as discussed in the text. While some peaks remain at the $< 0.1 \rm~ppt$ level, none of these reaches our required significance level. Panel (d) shows the amplitude spectrum from HIPPI-2 $Q/I$ polarimetry. The complex quasiperiodic structure visible is due to diurnal (and other) aliasing, and illustrates the value of using multiple data sets to determine which peaks are ``real''. Here the red and grey vertical lines mark the two detected oscillation modes in this data set, while the matching dotted lines show the locations of predicted $\pm \rm 1,2,3,...~d^{-1}$ alias peaks. Panel (e) shows the same HIPPI-2 $Q/I$ amplitude spectrum after prewhitening 23 frequencies. The $U/I$ spectra are similar and are shown in Supplementary Figure \ref{fig:ASU}. Note: 1 ppt = 1000 ppm.}
  \label{fig:AS}
\end{figure*}

Subsequently, we apply three restrictions from the polarimetry. As a first restriction we demand compliance with the allowed inclination angles for each of the options for the dominant mode with frequency $f_2$ (Figure\,\ref{fig:f2_phase}). This restricts the solution space to 98 options whose $i$ and $v\sin i$ are shown in the insets in Figure\,\ref{fig:inclvsini}. It can be seen that $v\sin i$ is very well constrained thanks to this condition imposed by the dominant mode in the polarimetry. As a second condition from polarimetry, we require that $f_6$ has a mode with a measurable $A^M_U$ ($>2$\,ppm) -- since we have a clear detection for this -- which whittles the viable options down to 30. Finally, we eliminate any option with a significant predicted amplitude that is not detected; for this condition we conservatively set the threshold at $5.5$\,ppm -- this value is chosen to be $0.5$\,ppm above the top of the noise floor seen in Figure\,\ref{fig:AS}(e), and just below the detected amplitude of $f_2$ in $U/I$. There are eight options that strictly meet these criteria, and another one that could when one considers nominal errors.

Together with the predicted polarimetric amplitudes, the final solutions are given in Supplementary Table \ref{tab:MI-best}, listed in order of highest probability in complying with all constraints from spectroscopy and polarimetry together. Notably, the mode with frequency $f_5$ must be a zonal mode ($m_5=0$) from its zero amplitude in the line center and its step function in phase\cite{Telting1997}, as shown in the lower right panel of Figure\,\ref{fig:lpa}. The best solution (A) identifies $f_2$ as a $(3,-3)$ mode and $f_1$ as $(1,-1)$, for an inclination angle of $46^\circ\pm 2^\circ$ and $v\sin i=16\pm 2\,$km\,s$^{-1}$. Furthermore, since $f_2$ and $f_8$ are found to belong to the same multiplet, $f_2-f_8=0.0431$\cpd, this will allow the deduction of the envelope rotation rate, at the position inside the star where these modes have their dominant probing power. A concrete value can only be derived from dedicated asteroseismic models\cite{Aerts2019}.

\subsection*{Asteroseismic models}

To get a first assessment of stellar models compliant with our mode identification, we apply a neural network trained on a large grid of stellar models covering masses from 2 to 20\,M$_\odot$ and treating zonal modes of degree 0, 1, 2\cite{HendriksAerts2019}. Since this neural network does not predict modes of higher degree, we fed it with the frequency $f_1$ as a dipole mode and assumed $f_5$ to be a quadrupole zonal mode, as found for the best solution (A) for the mode identification. Although $f_1$ is not zonal, the frequency shift induced by rotation is very small, of order 0.02\,d$^{-1}$ as deduced from $f_2$ and $f_8$. We explicitly checked that this frequency difference does not affect the results discussed below.

\begin{figure*}
\begin{center}
\rotatebox{270}{
\includegraphics[clip, trim={0cm, 0cm, 0cm, 0.cm}, width=12.4cm, bb = 0 0 8.26in 11.69in]{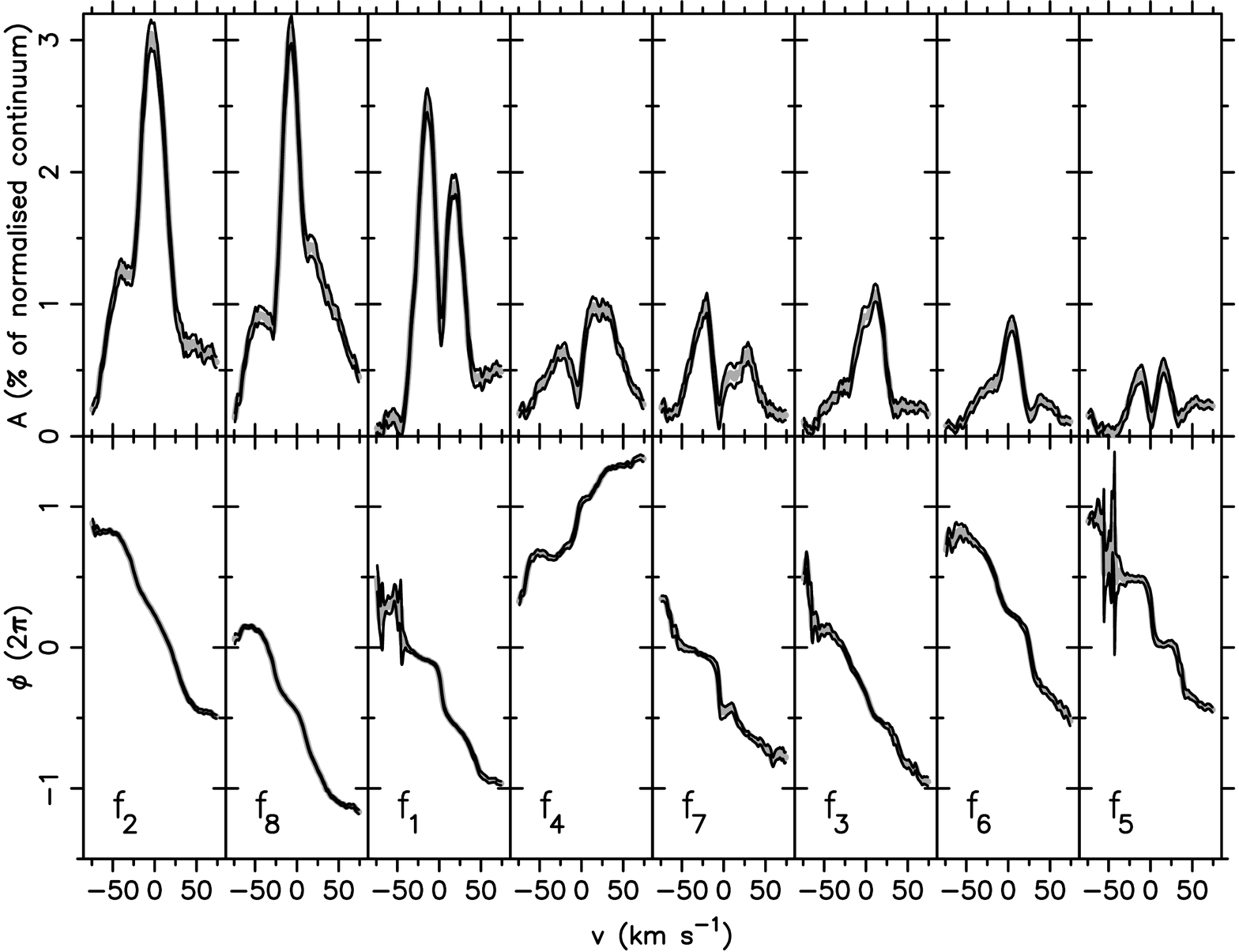}}
\end{center}
\caption{\textbf{Line profile analysis for $\beta$\,Cru |} The residual Si\,III
  line at 4552.6\,\AA\ from which the average profile was subtracted was fit with
  a linear multiperiodic oscillation model of the form
  $p_{\rm spec}(\lambda,t)=\sum_{i=1}^8 A_{i,\lambda} \sin (2\pi\ f_i\ t +
  \phi_{i,\lambda})$ for the eight frequencies labelled in
  Table\,\ref{tab:frequencies}. The frequencies are shown here, left to right, in order of dominance. The thick grey line indicates the best linear least-squares fit and the area indicated by the thin black lines represents the $1\sigma$ uncertainty region.
  Modes with an upwards phase slope are zonal
  ($m=0$) or retrograde ($m>0$), while those with a downward phase slope are zonal ($m=0$)
  or prograde ($m<0$).}
  \label{fig:lpa}
\end{figure*}

The neural network only allows us to use zonal modes of $l\leq 2$. Although some of the solutions in Supplementary Table \ref{tab:MI-best} list $f_3$ and $f_4$ as zonal modes, we do not wish to rely on their identification because these modes' amplitude distribution is not characteristic of such a mode, in comparison with the one for $f_5$ which is symmetrical and drops to zero value in the line center (Figure \ref{fig:lpa}). We thus rely on the most secure identification for the two zonal modes, i.e. $f_1$ and $f_5$. Without any further constraint, the neural network delivers model solutions for a star with too high a mass at the edge of the model grid. We therefore used additional constraints from spectroscopy and from the \textit{Hipparcos} parallax to place $\beta\,$Cru in the Hertzsprung-Russell diagram (HRD). The bolometric luminosity was computed from the spectroscopic $T_{\rm eff}$ and $\log\,g$\cite{Morel2008} and combined with an estimate for the reddening, $E(B-V)\simeq 0.03$\,mag, from the polarimetry (see the \nameref{sec:suppinf}) to derive the reddening corrected apparent bolometric magnitude\cite{Pedersen2020}, and subsequently determine the luminosity using the \textit{Hipparcos} parallax. The $1\sigma$ and $3\sigma$ results are graphically depicted in Supplementary Figure\,\ref{fig:HRD}, along with evolutionary tracks for masses 12, 14, 16, 18\,M$_\odot$, for various levels of initial chemical composition $(X,Z)$, convective core overshooting, $f_{\rm ov}$, in an exponentially decaying diffusive description, and envelope mixing $D_{\rm mix}$\cite{Moravveji2015}. 

We used the position of $\beta\,$Cru in the HRD to restrict the neural network, initially the still rather broad mass range $[9,18]\,$M$_\odot$ (grey dots in Supplementary Figure\,\ref{fig:NN}) and then 
to the acceptable mass range deduced from the $3\sigma$ position box in the HRD, i.e., $[12,16]\,$M$_\odot$ (blue dots). For both applications, we fixed the initial chemical composition to the observed values of the metallicity and helium mass fraction, $Z=0.014$ and $Y=0.268$ respectively \cite{Morel2008}, resulting in an initial hydrogen mass fraction of $X=0.718$. We fixed this initial chemical composition given the tight $(M,Z)$, $(f_{\rm ov},Z)$ and $(D_{\rm mix},Z)$ relations\cite{Moravveji2015}. The outcome in Supplementary Figure\,\ref{fig:NN} reveals that the best solution occurs at a mass between 14 and 15\,M$_\odot$. The evolutionary stage is well determined by the estimate of the hydrogen mass fraction in the convective core, $X_c\in [0.30,0.33]$. The internal mixing parameters are not constrained, which is not surprising given we only modelled two modes\cite{HendriksAerts2019}. Relying on $X_c\in [0.30,0.33]$ and allowing for the full ranges for the convective core overshooting and envelope mixing covered by the grid, we find the star to have an age between 9.7 and 12.8 million years and a convective core mass between 25\% and 32\% of its mass. This is in good agreement with the need for higher than standard core masses as derived from eclipsing binaries in this mass range\cite{Tkachenko2020}. The neural network solution places the star in between the two evolutionary tracks indicated in black lines in Supplementary Figure\,\ref{fig:HRD}. 

We find the radius of the star to fall in the range from $7.3$ to $8.9$\,R$_\odot$. This is slightly larger than implied by the angular diameter as measured by intensity interferometry\cite{HanburyBrown1974} if using the new \textit{Hipparcos} reduction\cite{vanLeeuwen07} for parallax -- $6.6\,\pm\,0.6$\,R$_\odot$ -- but in good agreement if using the original \textit{Hipparcos} parallax determination\cite{Hipparcos} -- $8.2\,\pm\,0.5$\,R$_\odot$. Both our mass and radius estimates are in agreement with earlier values based on multicolour photometry\cite{Hubrig2006}. The range for the radius, combined with the inclination and spectroscopic estimate of $v\sin i\simeq 16\,$km\,s$^{-1}$, lead to an equatorial rotation velocity of $\simeq 22\,$km\,s$^{-1}$ and a surface rotation period between 13 and 17\,d. The corresponding rotation frequency does not occur prominently in the \textit{TESS} data, i.e., we do not see any conclusive evidence of rotational modulation. Such slow surface rotation is quite common among $\beta\,$Cep pulsators \cite{Hubrig2006,Morel2008}. Various physical phenomena can explain an effective slow-down during the star's evolution \cite{Aerts2019},
amongst which is magnetic braking. In the case of $\beta\,$Cru, 
the presence of a magnetic field is inconclusive so far \cite{Hubrig2009}.
Additional constraints on the internal rotation can in principle be derived. This requires dedicated future asteroseismic modelling based on all identified modes. Methods to do so have not yet been developed for cases with dominant $\ell>2$ modes as found here.

\bigbreak

\textbf{Data Availability:} The new data that support the plots within this paper and other findings of this study are available from ViZieR at {\tt https://cdsarc.cds.unistra.fr/viz-bin/cat/J/other/NatAs/6.154}.

\bigbreak

\textbf{Code Availability:} Our polarimetric modelling code is based on the publicly available {\tt ATLAS9}, {\tt SYNSPEC} and {\tt VLIDORT} codes. Our modified version of {\tt SYNSPEC} is available on request. The spectroscopic mode identification was performed with the software package {\tt FAMIAS} available from {\tt  https://fys.kuleuven.be/ster/Software/famias/famias}, applied to the time-series spectroscopy available from {\tt https://fys.kuleuven.be/ster/Software/helas/helas}. 
The neural network is available from {\tt https://github.com/l-hendriks/asteroseismology-dnn}.
The joint frequency analysis was conducted using a custom {\tt MATLAB} package which is available on request.
  
\bigbreak

\textbf{Acknowledgements:} This research has made use of the SIMBAD database and VizieR catalogue access tool, operated at CDS, Strasbourg, France. This research has made use of NASA's Astrophysics Data System. We thank the Director of Siding Spring Observatory, A/Prof. Chris Lidman for his support of the HIPPI-2 project on the AAT. We thank Prof. Miroslav Filipovic for providing access to the Penrith Observatory. DVC would also like to thank Prof. Miroslav Filipovic and Prof. Bradley Carter for their support of his initially unfunded research on this project in the form of adjunct positions at WSU and USQ. We thank Nathan Cohen, Giulia Santucci and Darren Maybour for assisting with the observations. We thank Dr. Wm.\,Bruce Weaver for useful comments on the manuscript. Funding for the construction of HIPPI-2 was provided by UNSW through the Science Faculty Research Grants Program (JB). Part of the research leading to these results has received funding from the European Research Council (ERC) under the European Union’s Horizon 2020 research and innovation programme by means of an ERC AdG to CA (grant agreement N$^\circ$670519: MAMSIE). This research was supported in part by the National Science Foundation under Grant No. NSF PHY-1748958 (MGP). DLB acknowledges support from the \textit{TESS} Guest Investigator Program through award NNH17ZDA001N-TESS. 

\bigbreak

\textbf{Author contributions:} All authors contributed to the discussion and drafting of the final manuscript. DVC, DLB, CA, JB, DS, MGP, PDC, LK-C and ADH contributed to observing proposals and/or scheduling. DVC, JB, DLB, ADH, LK-C, FL and SPM carried out polarimetric observations. In addition the following authors made specific contributions to the work: DVC initiated the work, contributed the polarimetric data processing and analysis, calculations of, and comparisons to, the analytical model, investigated binary effects, the interstellar polarization and co-ordinated the observations and analysis. DLB carried out the frequency analysis and contributed analysis of asteroseismic data. CA analysed the spectroscopic data and carried out the associated mode identification, as well as the asteroseismic modelling. JB contributed the polarized radiative transfer modelling, investigated binary effects and aided with interpretation of the analytical model. SB computed theoretical stellar models and pulsation modes for the asteroseismic modelling and ran the neural network. MGP computed bolometric corrections and the luminosity of $\beta$\,Cru, based on its spectroscopic properties. DS helped facilitate the initial collaboration and provided valuable context for the work.
\bigbreak

\textbf{Competing interests:} The authors declare no competing interests.
\bigbreak

\textbf{arXiv version notes:} This is the accepted version of the manuscript updated to include the link to online data, and correct a minor referencing error found in copyediting. The published paper reference is: Cotton et al. (2022), Nature Astronomy, Vol 6, pages 154–164 (s41550-021-01531-9). Two explainer article are also available: Cotton \& Buzasi (2022), Nature Astronomy, Vol 6, pages 24–25 (s41550-021-01547-1); Baade (2022), Nature Astronomy, Vol 6, pages 20-21 (s41550-021-01545-3).
\bigbreak

{\footnotesize\bibliography{refs}}

\bigbreak


\clearpage
\begin{table*}
\begin{center}
\tabcolsep 5 pt
\resizebox{\textwidth}{!}{%
\begin{tabular}{lccccccccc}
\toprule
ID   & 2002 \textit{WIRE} & \textit{WIRE}     & \textit{TESS}     & {$\langle v\rangle$}                & {$\langle v^2\rangle$}               & {$Q/I$}           & {$U/I$}           & {\textit{Hipparcos}}       & {\textit{WIRE}+\textit{TESS}} \\
     & $f$ (d$^{-1}$)  & $f$ (d$^{-1}$) & $f$ (d$^{-1}$) & $f$ (d$^{-1}$)    &  $f$ (d$^{-1}$)    & $f$ (d$^{-1}$) & $f$ (d$^{-1}$) & $f$ (d$^{-1}$) & $f$ (d$^{-1}$) \\
     &                    & $A$ (ppm)         & $A$ (ppm)         & $A$ ($\rm km~s^{-1}$) & $A$ ($\rm km^2~s^{-2}$)& $A$ (ppm)         & $A$ (ppm)         & $A$ (ppm)         & $A$ (ppm)\\ 
\midrule
$f_1$  &  5.228(1) & 5.229(3) & 5.2298(6)& 5.2298(2) & 5.2298(2)&           &         & 5.22977(5)& 5.229776(1) \\
       &           & 3523(50) & 3639(16) & 0.69(4)   & 441(37)  &           &         & 7291(73)  & 3705(17) \\ 
\hdashline[1pt/5pt]
$f_2$  &  5.956(7) & 5.96(1)  & 5.964(5) & 5.9636(2) &          & 5.964(1)  & 5.964(2)&           & 5.96364(2) \\
       &           & 634(47)  & 231(10)  & 0.67(4)   &          & 7.08(91)  & 5.78(97)&           & 251(12) \\ 
\hdashline[1pt/5pt]
$f_3$  & 5.477(2)  & 5.478(6) & 5.4777(8)& 5.477(2)  & 5.4777(2)&           &         &           & 5.477750(3) \\ 
       &           & 2730(28) & 1641(10) & 0.47(5)   & 412(37)  &           &         &           & 1872(10) \\ 
\hdashline[1pt/5pt]
$f_4$      &           & 5.22(2)  & 5.216(9) & 5.21628(9)& 5.2162(2)&           &         &           & 5.2162(6) \\
       &           & 311(34)  & 144(10)  & 0.61(5)   & 448(34)  &           &         &           & 78(11) \\ 
\hdashline[1pt/5pt]
$f_5$       &           &          & 5.699(9) &           & 5.699(13)&           &         &           & 5.6977(3) \\
       &           &          & 119(10)  &           & 302(36)  &           &         &           & 141(10) \\ 
\hdashline[1pt/5pt]
$f_6$       &           &          & 7.661(8) &           & 7.6605(6)& 7.659(4)  &         &           & 7.6605(3) \\
       &           &          & 143(9)   &           & 255(38)  & 2.89(62)  &         &           & 99(9) \\
\hdashline[1pt/5pt]
$f_7$       &           &          & 6.36(1)  &           & 6.359(4) &           &         &           & 6.3596(6) \\
       &           &          & 89(9)    &           & 336(39)  &           &         &           & 57(9) \\ 
\hdashline[1pt/5pt]
$f_8$       &           &          & 5.92(1)  &           & 5.9229(1)&           &         &           & 5.9229(6) \\
       &           &          & 90(9)    &           & 534(36)  &           &         &           & 55(9) \\ 
\hdashline[1pt/5pt]
$f_9$     &           & 0.82(3)  & 0.819(7) &           &          &           &         &           & 0.81902(2) \\
       &           & 316(44)  & 177(10)  &           &          &           &         &           & 209(12) \\ 
\hdashline[1pt/5pt]
$f_{10}$  & 2.785(13) & 2.78(2)  & 2.78(1)  &           &          &           &         &           & 2.77786(3) \\
       &           & 236(31)  & 112(9)   &           &          &           &         &           & 131(10) \\ 
\hdashline[1pt/5pt]
$f_{11}$  & 3.527(13) & 3.529(17)&	3.53(1)  &           &          &           &         &           & 3.52863(3) \\
       &           & 368(34)  & 118(10)  &           &          &           &         &           & 122(10) \\ 
\midrule
$\Delta f$  & 0.057 & 0.057 &	0.039  &   0.00024       &  0.00024      &  0.0021         &  0.0021       &   0.00091        & 0.00014 \\


\bottomrule
\end{tabular}}
\caption{\textbf{Significant frequencies found for $\beta$\,Cru |} Frequencies shown are those for which a $\rm SNR>4$ detection was made in two or more independent data sets. The polarimetric data, $Q/I$ and $U/I$ are derived from HIPPI-2 observations. The spectroscopic data, giving $<\nu>$ and $<\nu^2>$, were acquired with ESO's CAT/CES instrumentation \cite{Aerts1998}. The rightmost column shows frequency and amplitude determinations based on the combination of the \textit{WIRE} and \textit{TESS} time series, which provides the longest temporal baseline and hence the most precise frequency determination; as those two missions both were well-separated in time and had different bandpasses, the amplitudes derived from the combined time series should be considered as illustrative only. Mode detections and frequencies from the original 2002 analysis\cite{Cuypers2002} of the \textit{WIRE} data are also shown for reference. Mode IDs for $f_1 - f_8$ reflect assignments previously occurring in the literature and/or used in our spectroscopic mode identification (see \nameref{sec:methods}), while new identifications $f_9 - f_{11}$ are assigned in order of increasing frequency. The last line of the table gives the nominal frequency resolution for each data set, estimated from each time series length $T$ as $\Delta f = 1 / T$; this is most meaningful for the photometric time series, which have high duty cycles.}.
\label{tab:frequencies}
\end{center}
\end{table*}

\clearpage
\begin{figure*}
\includegraphics[clip, trim={0cm, 0.5cm, 0cm, 0cm}, width=\textwidth, bb = 0 0 8in 9.95in]{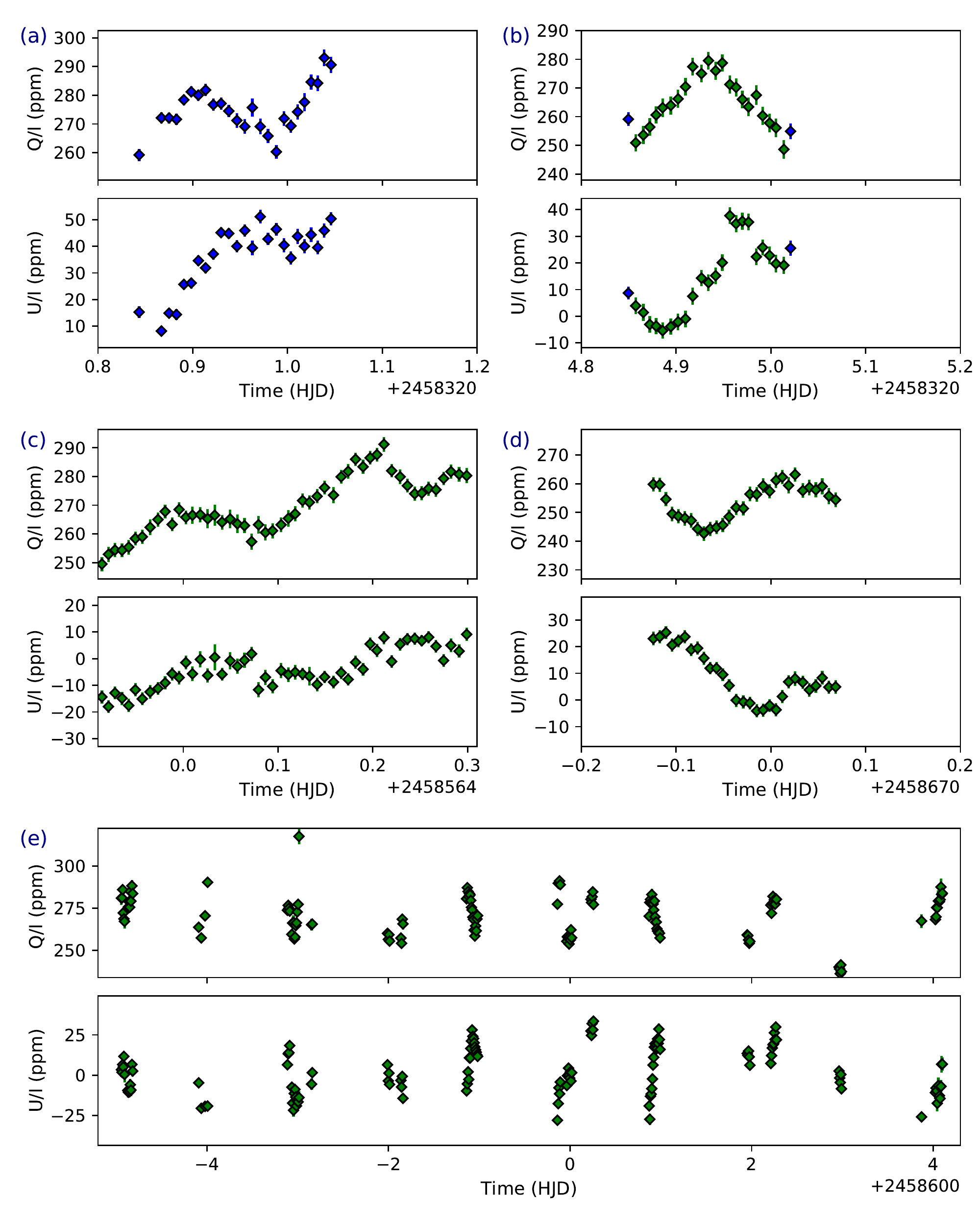}
\caption{\textbf{$\beta$ Cru time series examples |} Polarimetric observations of $\beta$ Cru are shown as pairs of time series for $Q/I$ (upper panels) and $U/I$ (lower panels) for four single nights (a -- d), and a 10-night run (e) to demonstrate the polarimetric variability on short time-scales. Unfiltered (Clear) observations are shown in blue, SDSS $g^{\prime}$ observations in green. The plotted uncertainties are the nominal 1-$\sigma$ errors, which are a combination of photon shot noise and an instrumental `positioning error' \cite{Bailey20}.
}
\label{fig:ts_examples}
\end{figure*}

\clearpage

\clearpage

\clearpage
\begin{figure*}
\begin{center}
\includegraphics[clip, trim={0.7cm, 0.3cm, 0.5cm, 0.1cm}, width=17.5cm, bb = 0 0 8.75in 8in]{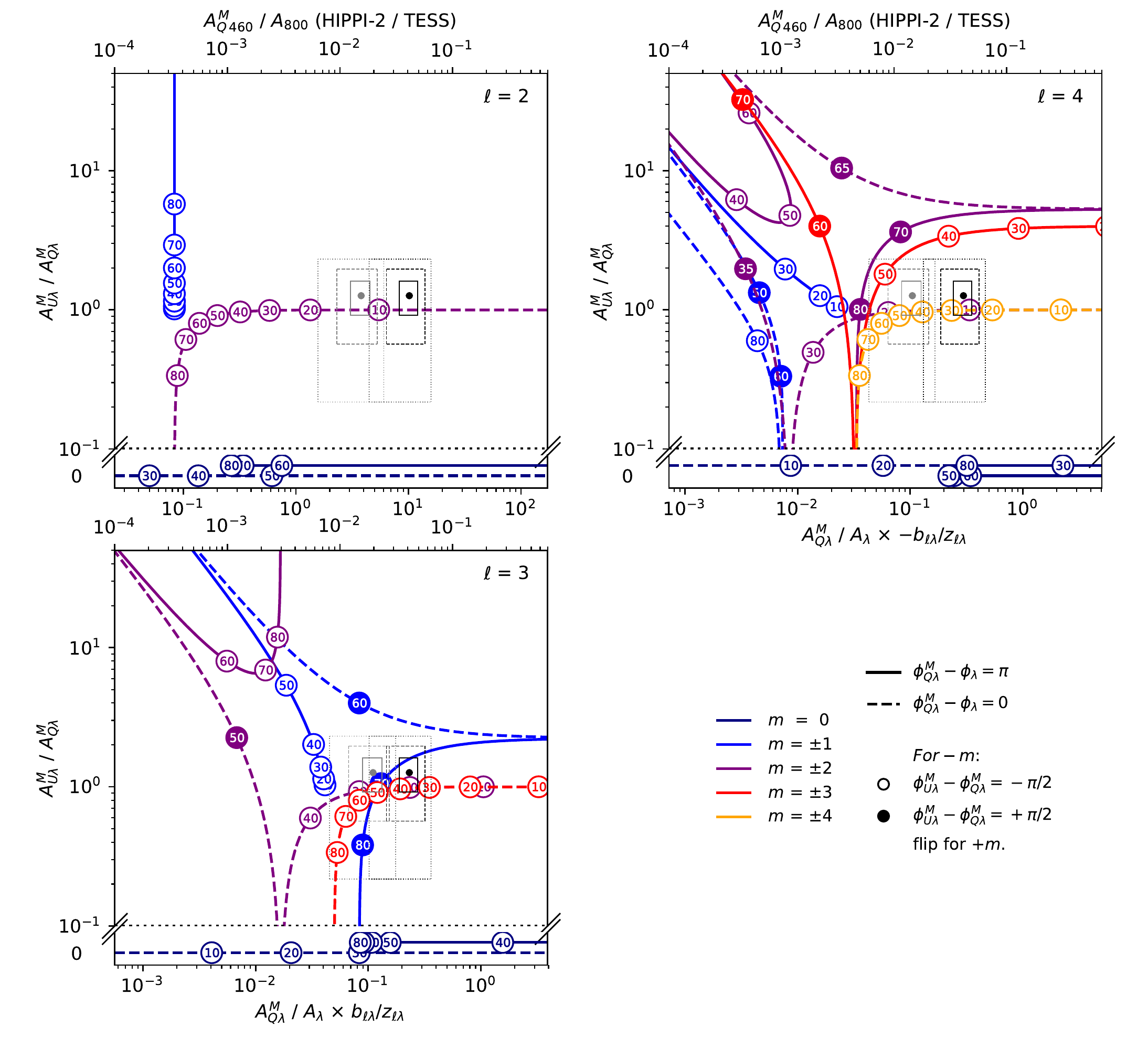}
\end{center}
\caption{\textbf{Polarimetric mode determination diagrams for $f_2$ mode of $\beta$\,Cru |} Allowed modes and observational data are shown in terms of the ratios of polarimetric amplitudes ($A_{U\lambda}^M/A_{Q\lambda}^M$) and polarimetric to photometric amplitudes ($A_{Q\lambda}^M/A_{\lambda}$, dimensionless units). Mapped onto the upper-left ($\ell=2$), lower-left ($\ell=3$) and upper-right ($\ell=4$) panels are the allowed $m$ values, where the numbers within the circles indicate stellar inclination. For $m=0$ note that $A_{U\lambda}^M/A_{Q\lambda}^M=0$ and the offsets in their display are for clarity only. The observational ratios for $f_2$ shown are those of HIPPI-2 $g^{\prime}$ (460~nm) polarimetric data and \textit{TESS} (800~nm, black dot) or \textit{WIRE} (600~nm, grey dot) photometric data; 1-, 2- and 3-$\sigma$ errors are represented as solid, dashed and dotted boxes, respectively. The data for $\beta$\,Cru $f_2$ shows $\phi_{U\lambda}^M-\phi_{Q\lambda}^M=-\pi/2$, meaning lines with open circles represent $-m$ modes and filled circles $+m$ modes. Similarly dashed lines correspond to $\Phi_\chi=25^\circ$ and solid lines $115^\circ$. The amplitudes are presented on the bottom axis in terms of $z_{\ell\lambda}$ and $b_{\ell\lambda}$ which are derived from a stellar atmosphere model (see \nameref{sec:methods}) and are dependent on $\ell$ and wavelength, $\lambda$, as well as $T_{\rm eff}$ and $\log g$. These values and their ratios for $\beta$\,Cru and the relevant wavelengths are given in Supplementary Table \ref{tab:zl_bl}. Note that when $z_{\ell\lambda}/b_{\ell\lambda}$ is negative, as it is for $\ell=4$, the solid and dashed lines are switched. The top axes show the amplitude ratios for $z_{\ell \: 460}$ and $b_{\ell \: 800}$.}
\label{fig:f2_mode}
\end{figure*}

\clearpage

\clearpage
\begin{figure*}
\begin{center}
\rotatebox{270}{
\includegraphics[width=9.5cm, bb = 0 0 8.26in 11.69in]{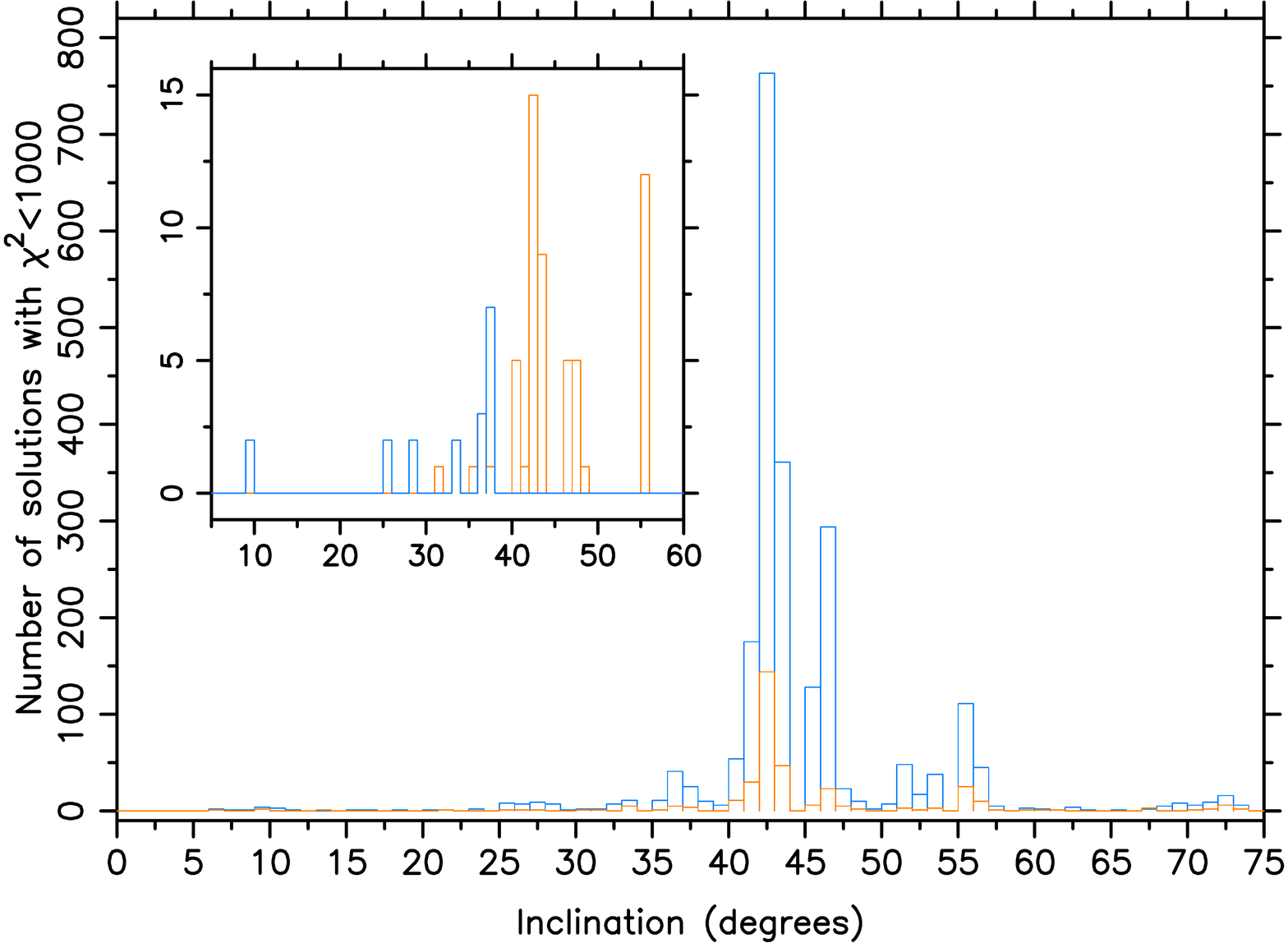}}
\rotatebox{270}{
\includegraphics[width=9.5cm, bb = 0 0 8.26in 11.69in]{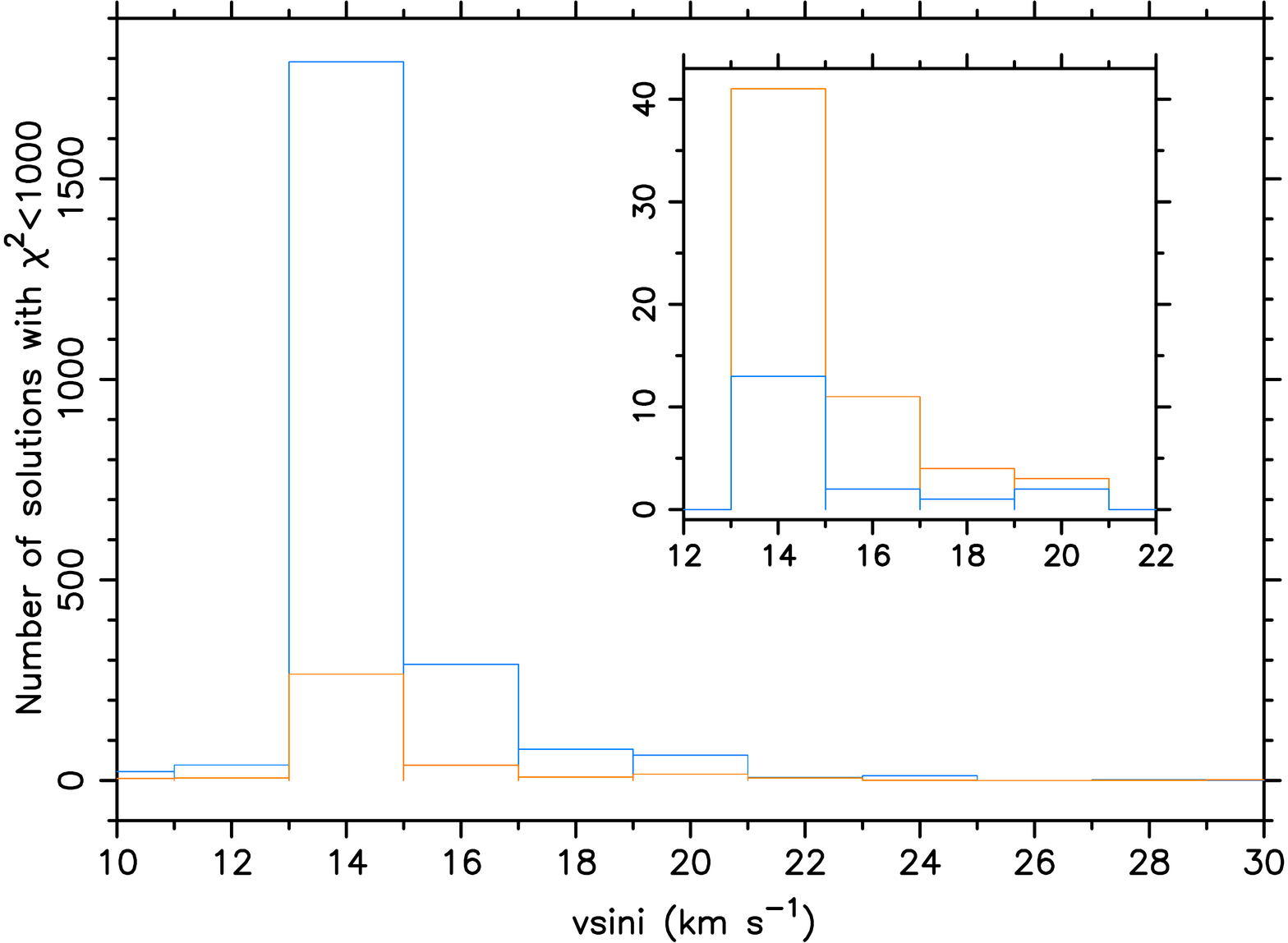}}
\end{center}
\caption{\textbf{Distributions of $\beta\,$Cru's rotation axis inclination angle |}
Histograms of the rotation inclination angle (upper panel) and of the projected surface rotation velocity (lower panel) for all the solutions of the line-profile fits having $\chi^2<1000$.
Among all considered cases of $\ell\in [0,4]$ for all the eight modes detected in the spectroscopy (cf.\,Fig.\,\ref{fig:lpa}), only solutions where the dominant mode in polarimetry has $\ell=3$ (orange) or $\ell=4$ (blue) fulfill $\chi^2<1000$ so we exclude a lower degree for that mode. The insets show the subsets of solutions when imposing the stringent constraints on the inclination angle derived from the polarimetry for each of the possibilities shown in Fig.\,\ref{fig:f2_mode}. This reveals the major gain in limiting the number of solutions from the inclination restrictions imported from the polarimetry.}
  \label{fig:inclvsini}
\end{figure*}


\clearpage
\newcommand{\hbAppendixPrefix}{S}
\renewcommand{\thefigure}{\hbAppendixPrefix\arabic{figure}}
\setcounter{figure}{0}
\renewcommand{\thetable}{\hbAppendixPrefix\arabic{table}} 
\setcounter{table}{0}
\renewcommand{\theequation}{\hbAppendixPrefix\arabic{equation}} 
\setcounter{equation}{0}

\newpage
\section*{Supplementary Information}
\label{sec:suppinf}
\phantomsection

\subsection{Photometric amplitude variability}

The ratio of $b_{\ell 600}$/$b_{\ell 800}$, which is calculated from the data in Supplementary Table \ref{tab:zl_bl}, describes the expected ratio of \textit{WIRE} to \textit{TESS} amplitudes. So, for $\ell=1$ and $\ell=3$ we expect a ratio of 1.01 and 1.31 respectively. Aside from the $\ell=1$ mode $f_1$, the actual amplitudes as given in Table \ref{tab:frequencies} are discrepant in comparison. The ratio for $f_1$ is 0.97, but for all of the other frequencies for which we have both \textit{WIRE} and \textit{TESS} amplitudes, the ratio falls between 1.66 and 3.19. This discrepant ratio is a clear indication that the amplitudes have changed with time. The \textit{WIRE} data were obtained $\sim$19\,years prior to the \textit{TESS} data and none of these data sets covers the entire beating cycle of all the detected modes, so this is reasonable. Variable amplitudes occur frequently in stars with heat-driven modes \cite{Pigulski2008}. As indicated in the main text, fitting of the frequency peaks in the power spectra of the various data sets with Lorentzian functions leads to widths of those peaks giving mode lifetimes that are  considerably longer than the data sets. This supports the conclusion we are dealing with coherent heat-driven modes excited by the $\kappa$ mechanism, which is the well-known excitation mechanism for low-order modes in the $\beta\,$Cep stars \cite{Pamyatnykh1999}.

\subsection[f4]{Resolving frequency $f_4$}
Based on the \textit{TESS} data alone, $f_1$ and $f_4$ are separated by $0.0138~\rm d^{-1}$, which is only 35\% of the nominal frequency resolution. The corresponding separation for the \textit{WIRE} data alone is $0.009~\rm d^{-1}$, which is 16\% of the nominal frequency resolution. It is therefore possible that these peaks are due to amplitude variability of a single mode rather than two closely-spaced mode frequencies beating against one another. We adopt the latter interpretation because (1) the methodology we use finds $f_4$ in the \textit{WIRE}, \textit{TESS}, and spectroscopic data sets, (2) while amplitude variability is common in space photometry of $\delta$ Sct stars, it does not frequently occur at the high level found here in $\beta$ Cep stars, and (3) the time scale of the amplitude change is too short to have a physical origin, while it is a natural time scale for the case of mode beating. Adopting this interpretation also fulfills the need to include a harmonic component with this frequency to achieve a proper regression fit to the observed time series data sets, and allows us to treat all data sets and frequencies in a consistent way.

As an additional check, we have also done extra time-series analyses for both \textit{WIRE} and \textit{TESS} light curves individually using the software package {\tt Period04}, both before and after applying a high-pass filter. In all of these cases, the frequency $f_4$ is recovered. Optimal interpretation of $f_4$ as either due to beating or to amplitude variability or both can be revisited in the future when 
uninterrupted data with longer time coverage becomes available. 

\subsection{Higher degree modes}

The amplitudes of modes of $\ell=2$ and $3$ have previously been calculated with the analytical model \cite{Watson83}, to which we have added $\ell=4$. For these, and additionally for $\ell=5$, we have calculated the scaling factors $z_{\ell\lambda}$ and $b_{\ell\lambda}$, the ratio of which, together with the geometrical terms, give the ratio of polarimetric to photometric amplitudes. In general the geometrical term will tend to be reduced as $\ell$ increases, since it describes the \textit{net} asymmetry over the whole stellar disc, however this is highly dependent on the specific mode geometry. It can be seen that this is true for $\ell=2$, $3$ and $4$ in Figure \ref{fig:f2_mode}.

The values tabulated in Supplementary Table \ref{tab:zl_bl} show $\left | z_{4\,\lambda}/b_{4\,\lambda} \right | \sim \left | z_{3\,\lambda}/b_{3\,\lambda} \right |$. In general the relative polarimetric detectability of $\ell=3$ and $4$ modes will, given equal photometric amplitudes, depend on the geometry. On the other hand $\left | z_{5\,\lambda}/b_{5\,\lambda} \right |$ is significantly larger, meaning $\ell=5$ modes are likely to be more detectable polarimetrically. 

As mentioned in the main text, modes with $\ell\geq5$ are not known to be prominent in space photometry of slowly rotating $\beta$\,Cep stars \cite{Briquet2009, Handler2009}. Part of the reason for this are the low $b_{5\,\lambda}$ values presented in Supplementary Table \ref{tab:zl_bl}. Consequently it is not likely that any of the frequencies we identified in Table \ref{tab:frequencies} are higher order modes. However, HIPPI-2 is very precise, so there might be modes below the photometric detection threshold that are still detectable with polarimetry.

With a threshold of 9 to 10\,ppm for frequency detection in the photometric data, all the frequencies nominally detectable in the polarimetric data alone may not be captured by the joint analysis. However, examination of Figure \ref{fig:AS} and Supplementary Figure \ref{fig:ASU} does not reveal any prominent unassigned high polarization amplitude modes. After the assigned frequencies and their aliases, the next most prominent peak appears to correspond to $f_9$, which sits below our polarimetric detection threshold. Regarding $f_9$, we note that there are a number of peaks below $1~\rm d^{-1}$ in both the \textit{TESS} and \textit{WIRE} amplitude spectra, and these could arise from either stellar or instrumental sources. We include $f_9$ in our list because it is the only low-frequency peak that appears in both photometric amplitude spectra at $SNR > 4$.

In conclusion, modes of order $\ell=5$ might be detectable, but their scarcity in photometry for this type of star, and an inspection of our data argues against their presence at dominant amplitudes.

\subsection[f6]{Assignment of \textit{f}$_6$ and a note on polarimetric non-detections}

Frequency $f_6$ was detected polarimetrically in just one Stokes parameter with an amplitude $A_Q = 2.89 \pm 0.62$. The detection means that $\ell_6>1$. The amplitude is very close to the detection limit, so a similar amplitude pulsation in $U$ is not ruled out. 

Having determined $\Phi_\chi = 25^\circ$ from our analysis of $f_2$, we first rotated the data by twice this angle to get it into the model frame. Then the data, prewhitened by removal of the frequencies of stronger amplitude, was refit as $p=A \sin (2\pi f t + \phi)$, locked to the \textit{TESS} frequency $f_6 = 7.6593$\cpd, in the same fashion as for $f_2$ (see \nameref{sec:methods}). 

The result, shown in Supplementary Figure \ref{fig:f6_phase}, gives $A_U^M = 2.32 \pm 0.61$\,ppm, $\phi_U^M - \phi=-1.755 \pm 0.277$\,rad (within error of $-\pi/2$). No convincing fit for $Q^M$ was found. However, from the line profile analysis we have determined $f_6$ to be prograde ($m<0$); therefore $\phi_U^M - \phi_Q^M = -\pi/2$, and hence $\phi_Q^M - \phi = 0$, which rules out $m=1$ solutions (from $i\sim46^\circ$ and Figure \ref{fig:f2_mode}). Solutions with $m=0$ were, incidentally, also ruled out by virtue of a non-zero $A_U^M$; this has significant implications for the allowed solutions to the spectroscopic mode identification scheme as described in the \nameref{sec:methods}.

\begin{figure*}
\begin{center}
\includegraphics[clip, trim={0cm, 0cm, 0cm, 0.cm}, width=\textwidth, bb = 0 0 9.11in 6in]{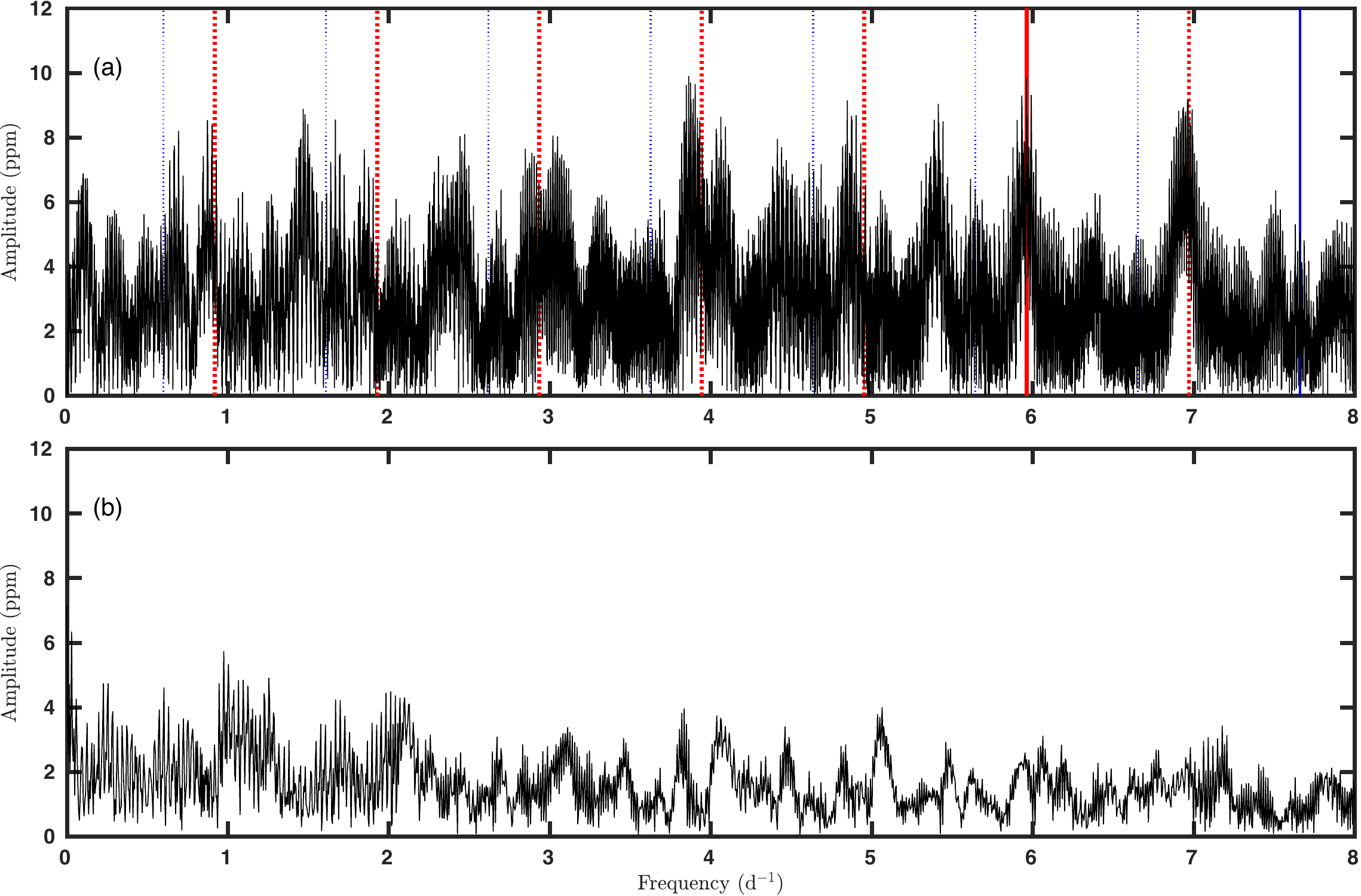}
\end{center}
\caption{\textbf{HIPPI-2 $U/I$ amplitude spectra for $\beta\,$ Cru |} The companion to Figure \ref{fig:AS} which shows photometric and $Q/I$ amplitude spectra. Here panel (a) shows the amplitude spectrum from HIPPI-2 $U/I$ polarimetry. The complex quasiperiodic structure visible is due to diurnal (and other) aliasing, and illustrates the value of using multiple data sets to determine which peaks are ``real''. Here the red vertical line marks the detected oscillation mode in this data set, and the blue vertical line the additional mode detected in $Q/I$, while the correspondingly coloured dotted lines show the locations of predicted $\pm \rm 1,2,3,...~d^{-1}$ alias peaks. Panel (b) shows the same HIPPI-2 $U/I$ amplitude spectrum after prewhitening 23 frequencies, as discussed in the text. Note: 1 ppt = 1000 ppm.}
  \label{fig:ASU}
\end{figure*}

Most of the remaining solutions given in Supplementary Table~\ref{tab:MI-best} produce predicted amplitudes for $f_6$ where $A_Q^M$ should be slightly larger than $A_U^M$, so we might have expected to detect it. However, at these very small polarizations, small errors in TP calibration between runs, PA imprecision, or an imperfect determination of the zero point (determined from the error weighted mean of the observations), could be enough to increase the noise or dampen the signal in one Stokes parameter to produce such a result. This is consistent with $f_2$ also showing $A_Q^M<A_U^M$, and also with the $U/I$ amplitude spectrum (Supplementary Figure~\ref{fig:ASU} being noisier than the $Q/I$ amplitude spectra (Figure~\ref{fig:AS}) -- since after rotation into the $^M$-frame $Q/I$ is mostly $U^M/I$ and vice-versa ($2\Phi_\chi=50^\circ$).

Most of the possible mode solutions for other frequencies given in Supplementary Table~\ref{tab:MI-best} also have $A_Q^M>A_U^M$, so the fact that the $U/I$ spectrum is noisier than $Q/I$ could also support the hypothesis that other modes below our detection threshold are impacting the observed polarimetric variability.

\subsection{Interstellar polarization and reddening}

The error weighted mean values of $Q/I=266.5\pm0.2$\,ppm and $U/I=3.1\pm0.2$\,ppm represent the constant polarization component of the star system unassociated with asteroseismic variability. Since $\beta$\,Cru is not a rapidly rotating star nor does it possess a debris disk, this constant value is assigned to the effect of aligned dust grains along the sight line to the star. Generally, such interstellar polarization is correlated with distance \citeS{Hiltner49,Draine03} but the relationship depends on the particular observed area on the sky as the dust density in the Galaxy is unevenly distributed.

$\beta$\,Cru is 88\,pc distant from the Sun \cite{vanLeeuwen07}. This places it near the ``wall'' of the Local Hot Bubble (LHB) -- a region of space with low gas and dust density that extends 75 to 150\,pc from the Sun. Within the LHB the interstellar medium polarizes at a rate of $\sim$0.2 to 2\,ppm\,pc$^{-1}$ \cite{Cotton16}$^,$\citeS{Cotton17b}, compared to tens of ppm\,pc$^{-1}$ beyond it \citeS{Behr59}. The LHB wall is a poorly defined transition region where the interstellar polarization rate increases sharply \citeS{Cotton19, Gontcharov19}. The constant polarization component we measure for $\beta$\,Cru, being somewhat larger than the trend expected within the LHB, is broadly consistent with being in the wall.

Efforts are underway to map the interstellar polarization at ppm sensitivity \cite{Cotton16}$^,$\citeS{Cotton17b, Piirola20} but they do not extend to the distance of $\beta$\,Cru with sufficient spatial density to be useful. In the agglomerated stellar polarimetry catalogue of Heiles \citeS{Heiles00} there are two stars with nominal errors $<$\,250\,ppm, within 5$^\circ$ of $\beta$\,Cru and between 60 and 110\,pc. The catalog data is given in terms of the magnitude of polarization, $P=\sqrt{(Q/I)^2+(U/I)^2}$ and the position angle of the direction of vibration, $\zeta=\frac{1}{2}\arctan{(\frac{U}{Q})}$. These two stars, both sourced by Heiles from the catalog of Reiz \& Franco\citeS{Reiz98}, are HD\,111102: $P=490\pm170$\,ppm, $\zeta=178\pm10^\circ$, which is not consistent with what we find for $\beta$\,Cru, and HD\,113152: $P=320\pm180$\,ppm, $\zeta=93\pm16^\circ$, which is consistent with what we find for $\beta$\,Cru. 

HD\,111102 is a double star where the primary is a B-type star \citeS{Fabricius02}, and so may be intrinsically polarized \cite{Cotton16}. Furthermore, the later catalog of Gontacharov \& Mosenkov \citeS{Gontcharov19} which recalculates the distances of earlier catalogues based on \textit{Gaia} data, places HD\,111102 further away at 114\,pc; consequently this star is not a good interstellar calibrator for $\beta$\,Cru. Gontacharov \& Mosenkov \citeS{Gontcharov19} find that HD\,113152 is also further away at 105\,pc, but it is still the best calibrator available and the agreement with the constant polarization component found for $\beta$\,Cru gives us confidence in attributing this to interstellar polarization.

The reddening of $\beta$\,Cru is poorly constrained by the current literature. Using the 2D reddening maps of Schlegel et al.\citeS{Schlegel98} gives $E(B-V)=0.92$\,mag -- which would be a remarkably high value for such a nearby star. A prescription by Bessel et al.\citeS{Bessel98}, that relies on the colour $(B-V)$ to estimated the reddening (cf. their Table~A5), gives a much lower, but still high, value of $E(B-V)=0.30$\,mag. This reddening value was previously used by Morel et al.\citeS{Morel08} to estimate the luminosity of $\beta$\,Cru. Using more recent 3D reddening maps \citeS{Marshall06, Gontcharov12} produces an even lower value of $E(B-V)=0.12$\,mag, but with quite large uncertainties. This wide range of interstellar reddening values results in differences in the estimated $\log L/L_\odot$ of up to 1\,dex.

Polarimetry is an alternative to standard methods of determining reddening that is likely to be more useful at short distances from the Sun. The maximum interstellar polarization, $p_{max}$, can be used to estimate reddening in the form of $E(B-V)$ \cite{Clarke10}: values for $p_{max}(\%) / E(B-V)$ have been seen to have extremes of $0.18$ and $9.0$\,\%$/$mag. However, the smallest value is attributed to Cassiopeia; for nearby stars a more applicable minimum of $1.0$ can be found by examining the results of Serkowski, Matthewson \& Ford \citeS{Serkowski75}. The determination of the interstellar polarization for $\beta$\,Cru here is made for a wavelength of $\sim460\,$nm. If the star is in the wall of the LHB, then its interstellar polarization will be a maximum at $550\,$nm \citeS{Cotton19}. Applying the Serkowski Law \citeS{Serkowski75, Wilking80, Whittet92}, we estimate $p_{max}$ to be $275\,$ppm. Consequently, we find that for $\beta\,$Cru $E(B-V) \leq 0.03$\,mag.

\subsection[bet Cru companions]{$\beta$\,Cru's companions}

The B0.5\,III type star $\beta$\,Cru A -- that we are concerned with in this paper -- has two companions, the second of which, $\beta$\,Cru\,D\footnote{The designation $\beta$\,Cru\,C has historically been used for a number of candidate companions that appear to have fallen out of favour.}, is 4$^{\prime\prime}$ distant and was discovered using the X-ray spacecraft \textit{Chandra} \citeS{Cohen08}. It is thought to be an active late-type pre-main sequence star \citeS{Cohen08}. Such a star might be expected to have a variable polarization associated with its activity that would add noise to our observations \citeS{Reiners12, Cotton17b, Cotton19b}. However, owing to $\beta$\,Cru\,D's tiny contribution to the total optical light this will not be significant. Variability in polarimetry associated with stellar variability of this companion, if we were to see any, is not as smooth as that expected for pulsations of the primary. We would expect much noisier data over the half night and longer runs shown in Figure \ref{fig:ts_examples} if the active companion were contributing to the polarimetric signal.

The first companion has already been mentioned in the \nameref{sec:methods}: $\beta$\,Cru\,B has a spectral type of B2\,V, an estimated mass below 10\,$M_\odot$ assuming aligned rotation inclination angles and contributes 8.5\% of the light at 443\,nm \cite{Aerts1998}$^,$\citeS{Popper68}. This will have the effect of dampening the measured polarization by a similar amount. However, the effect on polarimetric mode identification is robust to this because we only rely on amplitude \textit{ratios}. The ratio $A^M_{U\lambda}/A^M_{Q\lambda}$ is completely unaffected. However because the polarimetry is measured at a bluer wavelength than the photometry, the ratio $A^M_{Q\lambda}/A_{\lambda}$ is increased $\lesssim$\,5\% over its true value (based on simple blackbody calculations). In Figure \ref{fig:f2_mode} this ratio is plotted on a log scale, so it is evident that it would take a much larger discrepancy to influence the mode identification.

The orbit of $\beta$\,Cru\,B has a period of 1828\,$\pm$\,3\,d and an eccentricity of 0.38\,$\pm$\,0.09 \cite{Aerts1998}. Polarization can arise as a result of photospheric reflection in detached binary systems \cite{Bailey19}$^,$\citeS{Cotton20c}, but even allowing for the eccentricity, for a detectable signal the period would need to be on the order of days to weeks rather than years. 

There is no evidence $\beta$\,Cru\,B is also a pulsating star, but if it is, we expect any impact from this on our findings to be negligible. Given its much lower flux, the amplitudes of any oscillations would at minimum be proportionally smaller than those of the primary, given that $\beta$\,Cru\,A is a notably large amplitude pulsator. Additionally the secondary's lower mass implies that any dominant modes would likely be low-order pressure or gravity modes, which will tend to occur at a lower frequency \citeS{Szewczuk17, Aerts-CoRoT-2019}. We note that $f_{10}$ and $f_{11}$ are in this range, but are not used for mode identification, since they are not observed in either the spectroscopic or polarimetric data.

\bigbreak

\bibliographystyleS{naturemag}
{\footnotesize\bibliographyS{refs}}

\clearpage
\begin{figure*}
\includegraphics[clip, trim={0cm, 0cm, 0cm, 0cm}, width=\textwidth, bb = 0 0 8in 8in]{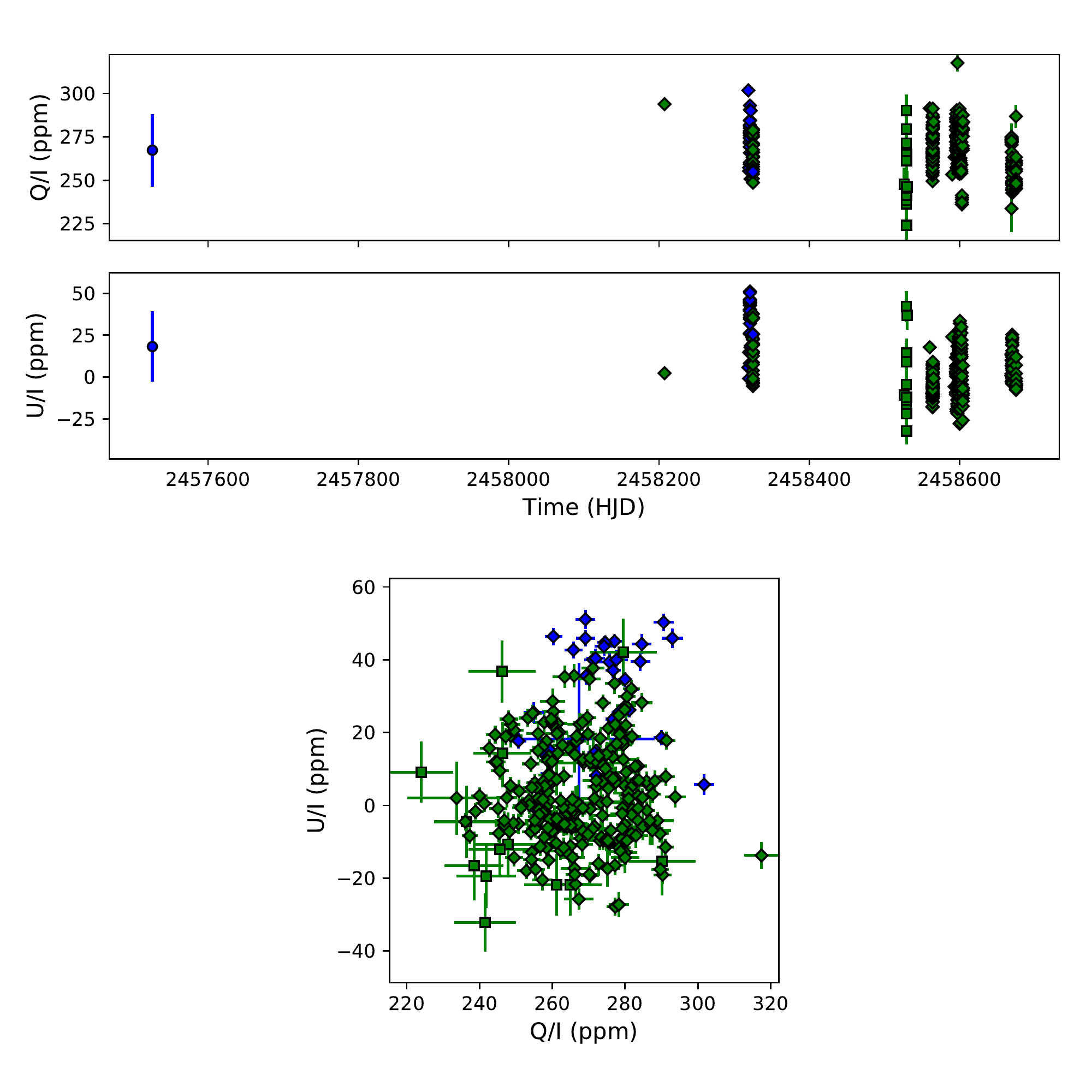}
\caption{\textbf{$\beta$ Cru observations |} All polarimetric observations of $\beta$ Cru (after calibration, subtraction of TP and efficiency correction) in time series for $Q/I$ Stokes (top panel), $U/I$ Stokes (middle panel), and as a Q-U diagram (bottom panel). Clear observations are shown in blue, SDSS $g^{\prime}$ in green, UNSW datum is a circle, WSU data are squares and AAT data are diamonds. Note that the errors associated with AAT observations are typically smaller than or on the same scale as the data points. The plotted uncertainties are the nominal 1-$\sigma$ errors, which are a combination of photon shot noise and an instrumental `positioning error' \cite{Bailey20}.}
\label{fig:observations}
\end{figure*}

\clearpage

\clearpage
\begin{figure*}
\begin{center}
\includegraphics[clip, trim={0cm, 0cm, 7.cm, 0cm}, width=12cm]{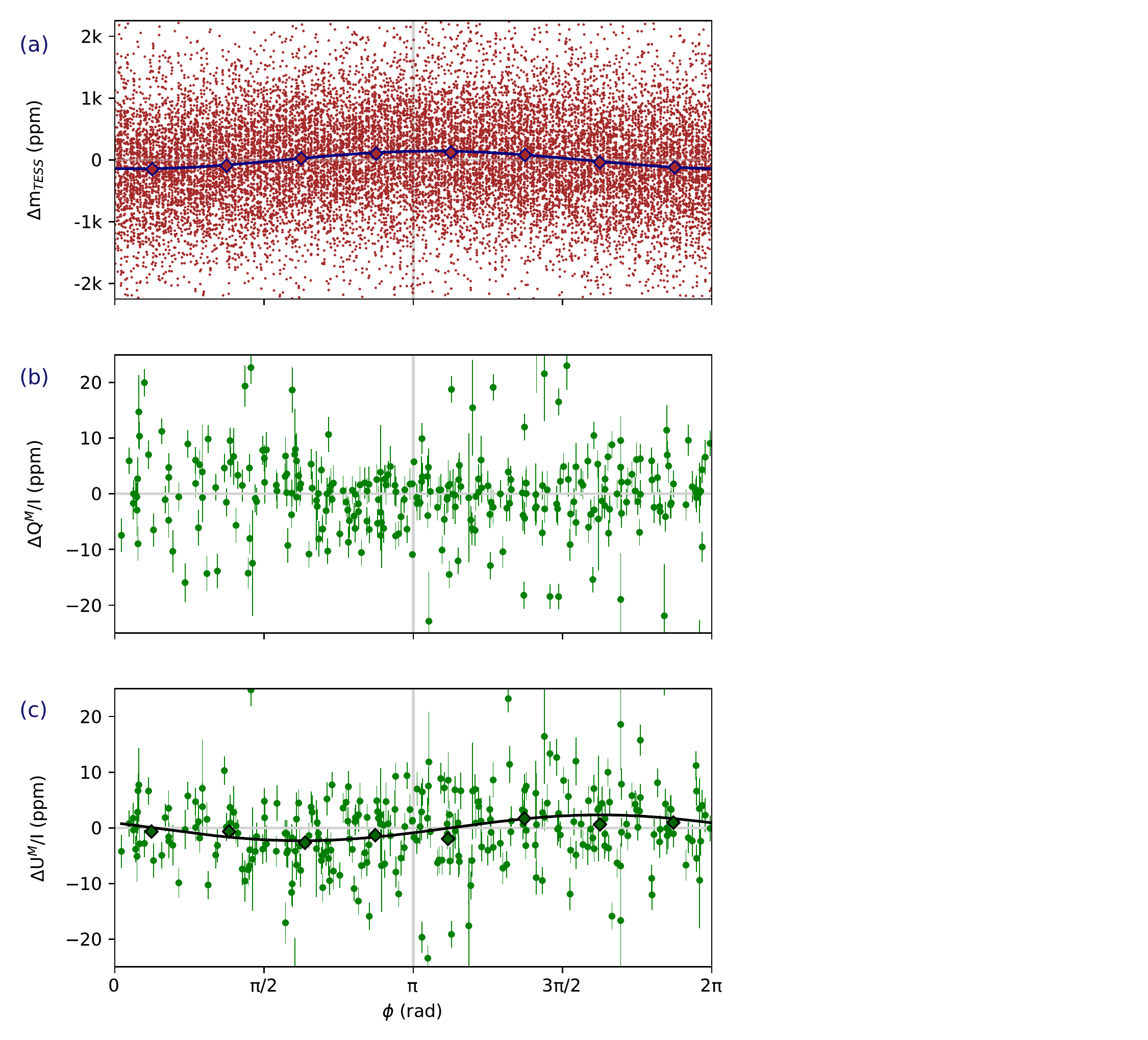}
\end{center}
\caption{\textbf{Photometric and polarimetric data phase-folded to $f_6$ for $\beta$\,Cru |} In panels (b) and (c) the polarimetric data, rotated 25$^\circ$ (to correspond to the $^M$ frame), is shown as the change in $Q^M/I$ and $U^M/I$ from the error weighted means as green dots (the plotted uncertainties are the nominal 1-$\sigma$ errors, which are a combination of photon shot noise and an instrumental `positioning error' \cite{Bailey20}). Note that the rotation effectively swaps the polarization between Stokes parameters. Similarly, photometric data from \textit{TESS} is shown as brown dots in panel (a). The data have been prewhitened by $f_2$. The sinusoidal fit to $U/I$ and photometric data, locked at the \textit{TESS} frequency $f_6 = 7.6593$\,d$^{-1}$, are shown as the solid black and blue lines, respectively. The diamonds in each plot represent the data binned in increments of $\pi/4$\,rad.}
\label{fig:f6_phase}
\end{figure*}

\clearpage
\begin{figure*}
\begin{center}
\rotatebox{270}{
\includegraphics[clip, trim={0cm, 0cm, 0cm, 0.cm}, width=12.5cm, bb = 0 0 7.17in 7.12in]{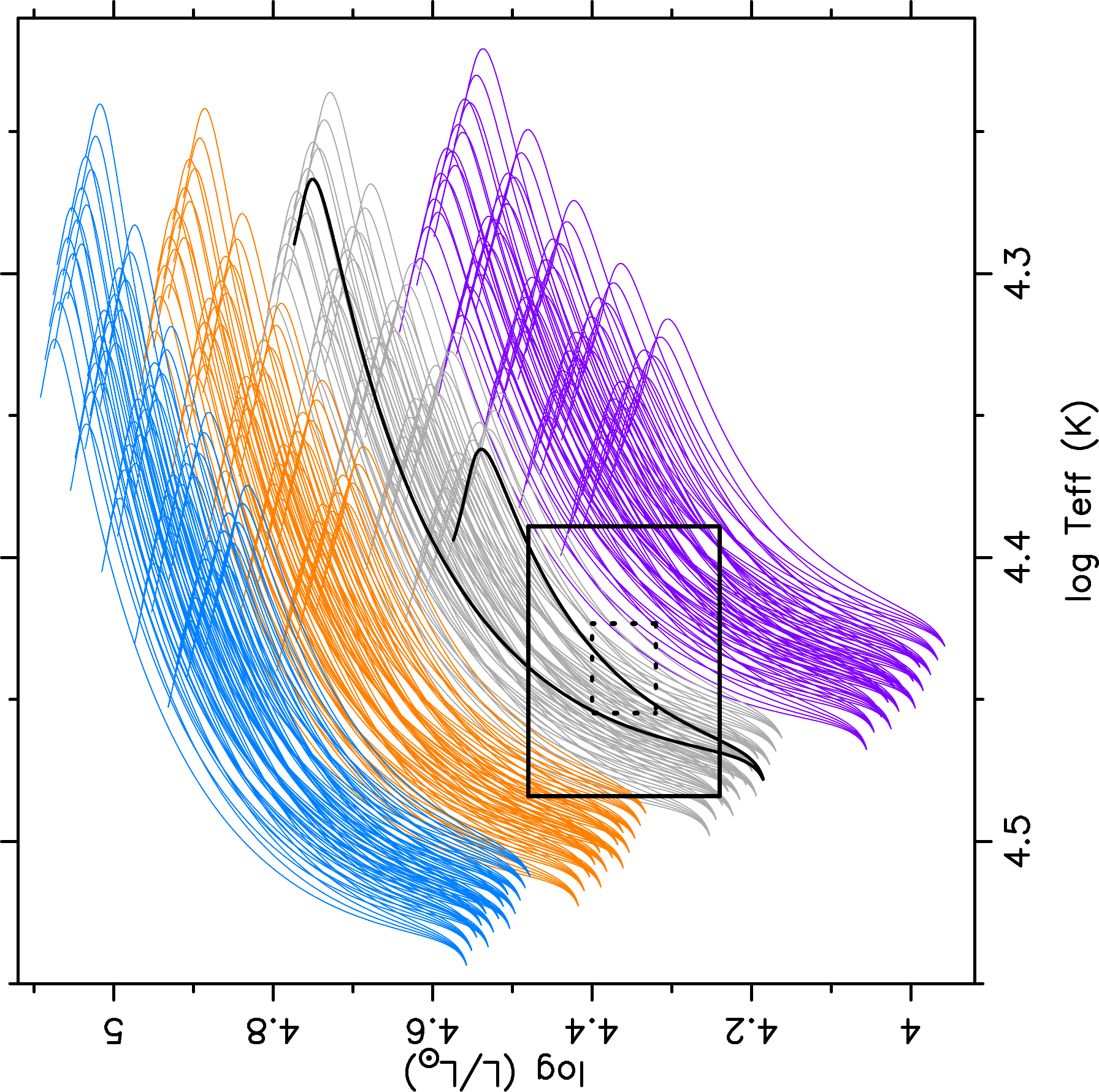}} 
\end{center}
\caption{\textbf{Position of $\beta\,$Cru in the Hertzsprung-Russell diagram |} 
Evolutionary tracks are shown for four masses (12 (purple), 14 (grey), 16 (orange), 18 (blue)\,M$_\odot$, five values of convective core overshooting $f_{\rm ov}$ (0.01, 0.02, 0.03, 0.04, 0.05 expressed in local pressure scale height), five values of the metallicity $Z$ (0.012, 0.014, 0.016, 0.018, 0.020), three values of the initial hydrogen mass fraction $X$ (0.68, 0.70, 0.72), and four values of the envelope mixing $D_{\rm mix}$ (1, 10, 100, 1000\,cm$^2$\,s$^{-1}$)\cite{HendriksAerts2019}. The position of $\beta\,$Cru is based on the spectroscopic temperature, our estimate of $E(B-V)$ from the polarimetry and a parallax measurement\cite{Pedersen2020} (dotted square: $1\sigma$; full square: $3\sigma$). 
Asteroseismic modelling based on the
two lowest-degree modes identified as dipole ($f_1$) and quadrupole ($f_5$) from application of a neural network (cf.\ Supplementary Figure\,\ref{fig:NN}) places $\beta\,$Cru 
in the middle of the core hydrogen burning stage between the two tracks indicated in black.}
  \label{fig:HRD}
\end{figure*}

\clearpage
\begin{figure*}
\begin{center}
\rotatebox{270}{
\includegraphics[clip, trim={0cm, 0cm, 0cm, 0.cm}, width=12.5cm, bb = 0 0 8.27in 11.69in]{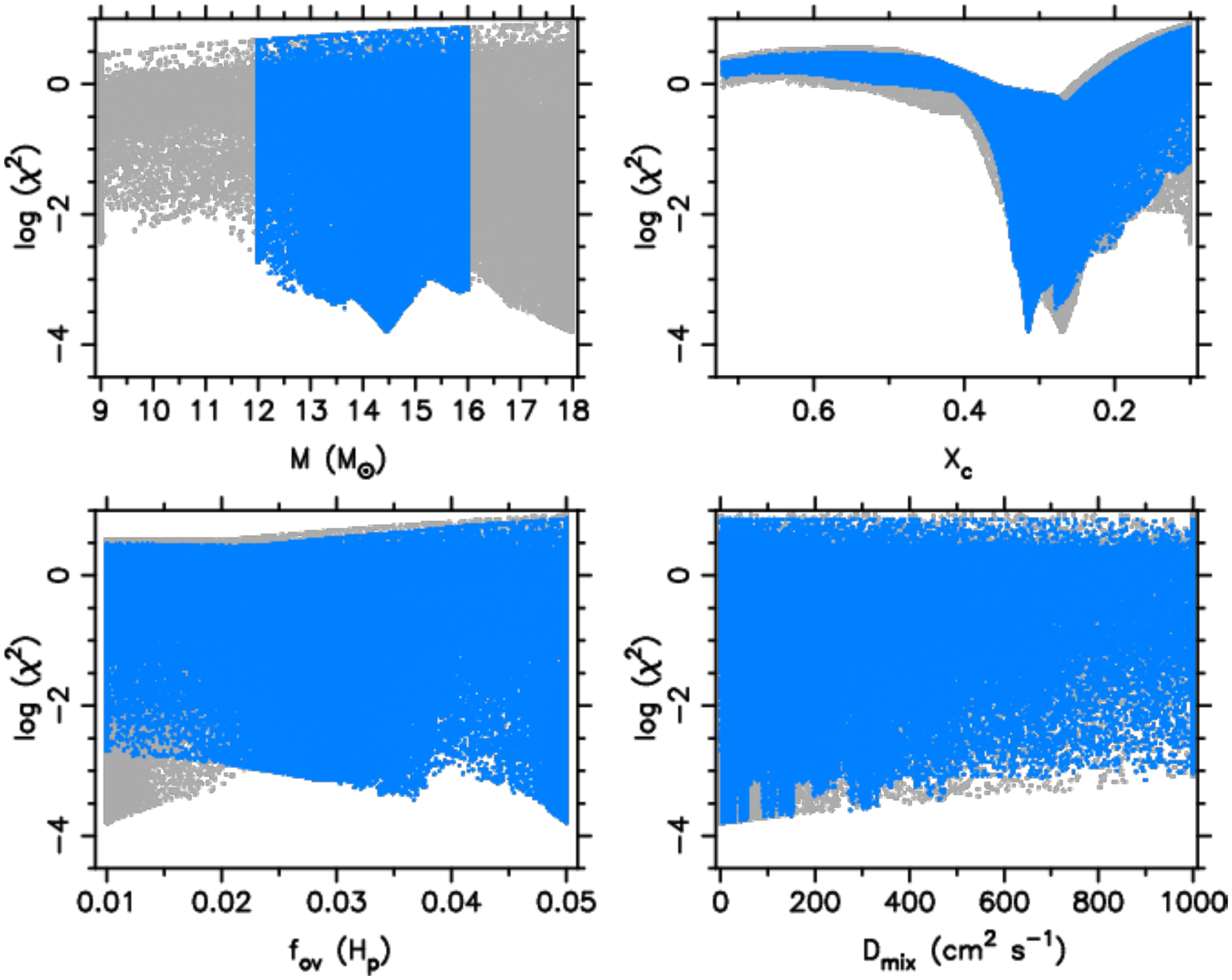}} 
\end{center}
\caption{\textbf{Asteroseismic modelling of $\beta\,$Cru from a neural network approach based on two zonal modes |} 
Estimation of the mass $M$, central hydrogen fraction $X_c$, convective core overshooting $f_{\rm ov}$ expressed in units of the local pressure scale height H$_{\rm p}$ in an exponentially decaying diffusive mixing approximation, and level of mixing in the envelope $D_{\rm mix}$ from application of a neural network trained on the grid of models for masses in the range $[9,18]\,$M$_\odot$ (gray) and restricted to the 3$\sigma$ error box of $\beta\,$Cru shown in Supplementary Figure\,\ref{fig:HRD} (blue); modes of degree $\ell=1$ and $\ell=2$ were imposed for the frequencies $f_1$ and $f_5$, respectively.}
  \label{fig:NN}
\end{figure*}

\clearpage
\begin{table*}
\begin{center}
\tabcolsep 3 pt
\resizebox{\textwidth}{!}{%
\begin{tabular}{cl|ccrccc|rrrrr}
\toprule
\multicolumn{2}{c|}{Run} & \multicolumn{6}{c|}{Telescope and Instrument Set-Up$^a$}   &   \multicolumn{5}{c}{Observations$^b$}    \\
ID & \multicolumn{1}{c|}{Date Range}   &   Instr.  &   Tel.   &   \multicolumn{1}{c}{f/}   &   Fil.  &   Mod.   &   Ap.     &   \multicolumn{1}{c}{n}   &   \multicolumn{1}{c}{Exp.}    &  \multicolumn{1}{c}{Dwell}   &   \multicolumn{1}{c}{$\lambda_{\rm eff}$}    &  \multicolumn{1}{c}{Eff} \\
& \multicolumn{1}{c|}{(UT)}  &     &      &     &    &      &    ($''$)    &      &   \multicolumn{1}{c}{(s)}    &   \multicolumn{1}{c}{(s)}   &    \multicolumn{1}{c}{(nm)}    &  \multicolumn{1}{c}{($\%$)} \\
\midrule
U1C & 2016-05-27  &   Mini-HIPPI  &   UNSW    &   11\phantom{.0*} &   Clear   &   MT    &  58.9    &   1   &   400 &   1284\phantom{$\pm$000} &   451.2\phantom{$\pm$0.0}  &   73.1\phantom{$\pm$0.0}   \\ 
A1G & 2018-03-29  &   HIPPI-2 &   AAT &   8*\phantom{.0} & $g^{\prime}$ &   BNS-E3  &   15.7    &   1 &   320 &   638\phantom{$\pm$000}   &   460.6\phantom{$\pm$0.0}   &   79.7\phantom{$\pm$0.0}   \\
A2C & 2018-07-19 to 25    &   HIPPI-2 &   AAT &   8*\phantom{.0} & Clear &   BNS-E4  &   11.9    &   33   &   320 &   619$\pm$\phantom{0}86 &   459.9$\pm$2.0    &   69.7$\pm$0.9   \\
A2G & 2018-07-25  &   HIPPI-2 &   AAT &   8*\phantom{.0} & $g^{\prime}$ &   BNS-E4  &    11.9    &   22   &   320 &   590$\pm$\phantom{0}39 &   459.7$\pm$0.6    &   78.3$\pm$0.3   \\
W1G & 2019-02-11 to 15    &   HIPPI-2 &   WSU &   10.5* &   $g^{\prime}$    &   ML-E1  &   58.9   &   14  &   480 &    1224$\pm$182    &   459.2$\pm$0.7  &   93.0$\pm$0.0    \\
A3G & 2019-03-17 to 21 &  HIPPI-2 &   AAT &   15\phantom{.0*}    &   $g^{\prime}$    &   ML-E1  &   12.7   &   54 &   320 & 618$\pm$\phantom{0}66  &   458.1$\pm$0.4   &   92.9$\pm$0.1  \\
A4G & 2019-04-16 to 30 &  HIPPI-2 &   AAT &   15\phantom{.0*}    &   $g^{\prime}$    &   ML-E1  &   12.7   &   128    &   320 & 581$\pm$112 &   458.1$\pm$0.5    &   92.9$\pm$0.0  \\
A5G & 2019-07-04 to 10 &  HIPPI-2 &   AAT &   15\phantom{.0*}    &   $g^{\prime}$    &   ML-E1  &   12.7   &   55 &   320 & 559$\pm$\phantom{0}38 &   458.4$\pm$0.6    &   93.0$\pm$0.0  \\
\bottomrule
\end{tabular}}
\caption{\textbf{Summary of polarimetric observations of $\beta$ Cru |} \textbf{*} Indicates use of a 2$\times$ negative achromatic lens effectively making the foci f/16 and f/21 respectively. \textbf{$^a$} A full description along with transmission curves for all the components and modulation characterisation of each modulator in the specified performance era have been described by Bailey et al. \cite{Bailey20}. \textbf{$^b$} The dwell time (Dwell), effective wavelength ($\lambda_{\rm eff}$) and modulation efficiency (Eff.) for the set of observations are described as the median $\pm$ the standard deviation. }
\label{tab:observations}
\end{center}
\end{table*}

\clearpage
\begin{table*}
\begin{center}
\tabcolsep 8.3 pt
\begin{tabular}{c|cccccccc|rr}
\toprule
Run &   \multicolumn{8}{c|}{Standard Observations*} & \multicolumn{1}{c}{$Q/I \pm \Delta Q/I$} & \multicolumn{1}{c}{$U/I \pm \Delta U/I$}\\
& A & B & C & D & E & F & G & H &\multicolumn{1}{c}{(ppm)} & \multicolumn{1}{c}{(ppm)} \\
\midrule
U1C & 0 & 0 & 4 & 0 & 0 & 0 & 0 & 0 &$-$90.8 $\pm$ 3.9 & $-$1.1 $\pm$ 3.9 \\
A1G & 0 & 0 & 3 & 3 & 3 & 0 & 0 & 0 &  130.1 $\pm$ 0.9 &  3.9 $\pm$ 0.9 \\
A2C & 3 & 3 & 2 & 2 & 1 & 0 & 0 & 2 &$-$10.1 $\pm$ 1.1 &  3.8 $\pm$ 0.9 \\
A2G & 3 & 3 & 2 & 2 & 2 & 0 & 0 & 2 &$-$12.9 $\pm$ 1.1 &  4.1 $\pm$ 1.0 \\
W1G & 0 & 0 & 4 & 0 & 0 & 0 & 0 & 0 &   12.7 $\pm$ 1.7 & 20.2 $\pm$ 1.7 \\
A3G & 0 & 0 & 3 & 2 & 0 & 0 & 0 & 0 & $-$8.8 $\pm$ 0.9 &  3.5 $\pm$ 0.7 \\
A4G & 0 & 0 & 4 & 3 & 0 & 0 & 2 & 1 &$-$10.0 $\pm$ 0.6 & $-$2.3 $\pm$ 0.6 \\
A5G & 3 & 0 & 0 & 3 & 0 & 0 & 0 & 0 &$-$14.5 $\pm$ 1.1 &  5.0 $\pm$ 1.0 \\
\bottomrule
\end{tabular}
\caption{\textbf{Low polarization standard observations to determine telescope polarization (TP) |} Indicated is the number of observations of each standard in each run. \textbf{*} Low polarization standards are the same as given by Bailey et al. \cite{Bailey20} and are as follows: A: HD\,2151, B: HD\,10700, \mbox{C: HD\,48915}, D: HD\,102647, \mbox{E: HD\,102870}, F: HD\,127762, G: HD\,128620J, H: HD\,140573. The adopted TP is given in the two right hand columns.}
\label{tab:tp}
\end{center}
\end{table*}

\clearpage
\begin{table}
\begin{center}
\tabcolsep 8.5 pt
\begin{tabular}{c|cccccccccccc|r}
\toprule
Run &  \multicolumn{12}{c|}{Standard Observations*} & S.D.\\
   & A & B & C & D & E & F & G & H & I & J & K & L & \multicolumn{1}{|c}{($^{\circ}$)} \\
\midrule
U1C & 0 & 0 & 4 & 0 & 0 & 0 & 0 & 0 & 0 & 0 & 0 & 0 & 0.10 \\
A1G & 0 & 1 & 0 & 1 & 1 & 0 & 0 & 0 & 0 & 1 & 0 & 0 & 0.26 \\
A2C & 0 & 0 & 0 & 0 & 1 & 0 & 1 & 1 & 0 & 0 & 1 & 0 & 1.56 \\
A2G & 0 & 0 & 0 & 0 & 1 & 0 & 1 & 1 & 0 & 0 & 1 & 0 & 1.56 \\
W1G & 0 & 0 & 1 & 0 & 1 & 0 & 0 & 0 & 0 & 0 & 0 & 0 & 0.11 \\
A3G & 0 & 1 & 1 & 0 & 1 & 0 & 0 & 0 & 0 & 0 & 0 & 0 & 0.46 \\
A4G & 0 & 1 & 1 & 0 & 1 & 0 & 0 & 0 & 0 & 0 & 0 & 0 & 0.27 \\
A5G & 0 & 0 & 0 & 0 & 4 & 0 & 1 & 0 & 0 & 0 & 0 & 0 & 0.08 \\
\bottomrule
\end{tabular}
\caption{\textbf{High polarization standard observations to calibrate position angle |} Indicated is the number of observations of each standard in each run. \textbf{*} High polarization standards are the same as those described by Bailey et al. \cite{Bailey20} and are as follows: A: HD\,23512, B: HD\,80558, C: HD\,84810, D: HD\,111613, E: HD\,147084, F: HD\,149757, G: HD\,154445, H: HD\,160529, I: HD\,161056, J: HD\,187929, K: HD\,203532, L: HD\,210121. S.D. is the standard deviation of the differences between literature and measured values for the position angle.\\}
\label{tab:pa}
\end{center}
\end{table}

\clearpage
\begin{table}
\begin{center}
\tabcolsep 3. pt
\renewcommand{\arraystretch}{1.6}
\begin{tabular}{cccc}
\toprule
$(\ell,\phantom{-}m)$  & $C_{m,-2}^{\,\ell}(i)+C_{m,+2}^{\,\ell}(i)$   & $C_{m,-2}^{\,\ell}(i)-C_{m,+2}^{\,\ell}(i)$ & $C_{m,\,0}^{\,\ell}(i)$ \\
\midrule
$(2,-2)$    &   $1 - \frac{1}{2}\sin^2(i)$ & $\cos(i)$  &   $\sqrt{\frac{3}{8}}\sin^2(i)$   \\
\hdashline[1pt/5pt]
$(2,-1)$    &   $-\sin(i)\cos(i)$&  $\sin(i)$   &  $\sqrt{\frac{3}{8}}\sin(2i)$ \\
\hdashline[1pt/5pt]
$(2,\phantom{-}0)$   &   $\sqrt{\frac{3}{2}}\sin^2(i)$    &  $0$   &   $\frac{1}{2}\left (3\cos^2(i) - 1 \right)$  \\
\hdashline[1pt/5pt]
$(2,+1)$    &   $-\sin(i)\cos(i)$   &   $\sin(i)$   &$-\sqrt{\frac{3}{8}}\sin(2i)$ \\
\hdashline[1pt/5pt]
$(2,+2)$    &   $1 - \frac{1}{2}\sin^2(i)$  &   $-\cos(i)$  &   $\sqrt{\frac{3}{8}}\sin^2(i)$   \\
\midrule
$(3,-3)$    &  $\sqrt{\frac{3}{8}}\sin(i)\left (1+\cos^2(i) \right)$    &   $\sqrt{\frac{3}{2}}\sin(i)\cos(i)$  &   $48\sqrt{720}\sin^3(i)$  \\
\hdashline[1pt/5pt]
$(3,-2)$    &   $\frac{1}{2}\cos(i)\left (3\cos^2(i) - 1\right)$    &   $2\cos^2(i) - 1$  &  $8\sqrt{120}\cos(i)\sin^2(i)$    \\
\hdashline[1pt/5pt]
$(3,-1)$    &   $\sqrt{\frac{5}{8}}\sin(i)\left (1 - 3\cos^2(i)\right)$ &    $-\sqrt{\frac{5}{2}}\sin(i)\cos(i)$ &   $\frac{1}{8}\sqrt{12}\sin(i)\left (5\cos^2(i) - 1\right)$   \\
\hdashline[1pt/5pt]
$(3,\phantom{-}0)$  &   $\sqrt{\frac{15}{2}}\sin^2(i)\cos(i)$   &   $0$ &   $\frac{1}{2}\left (5\cos^3(i) - 3\cos(i)\right)$    \\
\hdashline[1pt/5pt]
$(3,+1)$    &   $\sqrt{\frac{5}{8}}\sin(i)\left (1 - 3\cos^2(i)\right)$ &   $\sqrt{\frac{5}{2}}\sin(i)\cos(i)$  &   $-\frac{3}{24}\sin(i)\left (5\cos^2(i) - 1\right)$\\
\hdashline[1pt/5pt]
$(3,+2)$    &   $\frac{1}{2}\cos(i)\left (3\cos^2(i) - 1\right)$    &   $1 - 2\cos^2(i)$ &   $15\sqrt{\frac{1}{120}}\cos(i)\sin^2(i)$    \\
\hdashline[1pt/5pt]
$(3,+3)$    &   $\sqrt{\frac{3}{8}}\sin(i)\left (1+\cos^2(i) \right)$   &   $-\sqrt{\frac{3}{2}}\sin(i)\cos(i)$ &   $-15\sqrt{\frac{1}{720}}\sin^3(i)$\\
\midrule
$(4,-4)$    &   $\frac{\sqrt{7}}{8}\sin^2(i)\left (\cos(2i)+3\right)$   &   $\frac{\sqrt{7}}{2}\sin^2(i)\cos(i)$    &   $\sqrt{70}\cos^4(i/2)\sin^4(i/2)$   \\
\hdashline[1pt/5pt]
$(4,-3)$    &   $\frac{\sqrt{7}}{2}\sin(i)\cos^3(i)$    &   $\frac{\sqrt{7}}{2}\sin(i) \left (3\cos(2i) + 1\right)$    &   $-2\sqrt{35}\cos^3(i/2)\sin^3(i/2)\sin^2(i/2)$  \renewcommand{\arraystretch}{3.5}\\
\hdashline[1pt/5pt]
$(4,-2)$    &   $\frac{1}{16}\left (4\cos(2i) + 7\cos(4i) + 5\right)$   &   $\frac{1}{8}\left (\cos(i) + 7\cos(3i)\right)$  &   \renewcommand{\arraystretch}{1.4}\begin{tabular}{@{}c@{}} $\sqrt{10}\cos^2(i/2)\sin^2(i/2)3\sin^4(i/2)\times$    \\ $\left (3\cos^4(i/2) - 8\cos^2(i/2)\sin^2(i/2)\right)$ \end{tabular}\renewcommand{\arraystretch}{3.5}\\
\hdashline[1pt/5pt]
$(4,-1)$    &   $\frac{1}{8\sqrt{2}}\left(2\sin(2i) - 7\sin(4i)\right)$ &    $-\frac{1}{8\sqrt{2}}\left (3\sin(i) + 7\sin(3i)\right)$ & \renewcommand{\arraystretch}{1.4}\begin{tabular}{@{}c@{}}$2\sqrt{5}\cos(i/2)\sin(i/2)\times$ \\ $ (\cos^6(i/2) - 6\cos^4(i/2)\sin^4(i/2)$ \\ $+ 6\cos^2(i/2)\sin^2(i/2) - \cos(i/2)\sin^6(i/2))$\end{tabular}\renewcommand{\arraystretch}{1.6} \\
\hdashline[1pt/5pt]
$(4,\phantom{-}0)$  &   $\frac{1}{8}\sqrt{\frac{5}{2}}\sin^2(i)\left (12\cos(i) + 11\cos(2i) + 25\right)$ &   $0$ &   $\frac{1}{8}\left (35\cos^4(i) - 30\cos^2(i) + 3\right)$   \\
\hdashline[1pt/5pt]
$(4,+1)$    &   $-\frac{1}{8\sqrt{2}}\left(2\sin(2i) - 7\sin(4i)\right)$    &   $-\frac{1}{8\sqrt{2}}\left (3\sin(i) + 7\sin(3i)\right)$ &  \renewcommand{\arraystretch}{1.4}\begin{tabular}{@{}c@{}}$-2\sqrt{5}\cos(i/2)\sin(i/2)\times$ \\ $ (\cos^6(i/2) - 6\cos^4(i/2)\sin^4(i/2)$ \\ $+ 6\cos^2(i/2)\sin^2(i/2) - \cos(i/2)\sin^6(i/2))$\end{tabular}\renewcommand{\arraystretch}{1.6}\\
\hdashline[1pt/5pt]
$(4,+2)$    &   $\frac{1}{16}\left (4\cos(2i) + 7\cos(4i) + 5\right)$   &   $-\frac{1}{8}\left (\cos(i) + 7\cos(3i)\right)$ &   \renewcommand{\arraystretch}{1.4}\begin{tabular}{@{}c@{}} $\sqrt{10}\cos^2(i/2)\sin^2(i/2)3\sin^4(i/2)\times$    \\ $\left (3\cos^4(i/2) - 8\cos^2(i/2)\sin^2(i/2)\right)$ \end{tabular}\renewcommand{\arraystretch}{1.6}  \\
\hdashline[1pt/5pt]
$(4,+3)$    &   $-\frac{\sqrt{7}}{2}\sin(i)\cos^3(i)$   &   $\frac{\sqrt{7}}{2}\sin(i) \left (3\cos(2i) + 1\right)$    &   $2\sqrt{35}\cos^3(i/2)\sin^3(i/2)\sin^2(i/2)$\\
\hdashline[1pt/5pt]
$(4,+4)$    &   $\frac{\sqrt{7}}{8}\sin^2(i)\left (\cos(2i)+3\right)$   &   $-\frac{\sqrt{7}}{2}\sin^2(i)\cos(i)$   &   $\sqrt{70}\cos^4(i/2)\sin^4(i/2)$\\
\bottomrule
\end{tabular}
\caption{\textbf{Angular momentum transformation matrix elements and combinations |} These terms are needed to calculate the polarimetric and photometric amplitudes according to Watson's\cite{Watson83} Equations (15), (16) and (17). Many of the terms are given by Watson \cite{Watson83}, for the remainder we have used the same source tables as him to calculate them -- those of Buckmaster }
\label{tab:matrix-elements}
\end{center} {\citeS{Buckmaster64}}. 
\end{table}

\clearpage
\begin{table}
\begin{center}
\tabcolsep 8.5 pt
\begin{tabular}{c|c|cc|cc}
\toprule
$\ell$  &   \multicolumn{1}{c}{$z_{\ell\lambda}$}   &   \multicolumn{2}{|c}{$b_{\ell\lambda}$}   &   \multicolumn{2}{|c}{$z_{\ell\lambda}/b_{\ell\lambda}$} \\
        &   460\,nm &   600\,nm &   800\,nm     & 460\,nm\,/\,600\,nm   &   460\,nm\,/\,800\,nm \\
\midrule
    0   &   0.\0\0\0\0\0    &  \phantom{$+$}1.\0\0\0\0\0    & \phantom{$+$}1.\0\0\0\0\0     &   \phantom{$+$}0.\0\0\0\0\0    &  \phantom{$+$}0.\0\0\0\0\0 \\
    1   &   0.\0\0\0\0\0    &  \phantom{$+$}0.6842\0        &  \phantom{$+$}0.6802\0        &   \phantom{$+$}0.\0\0\0\0\0    &  \phantom{$+$}0.\0\0\0\0\0 \\
    2   &   0.00111         &  \phantom{$+$}0.2802\0        &  \phantom{$+$}0.2734\0        &   \phantom{$+$}0.00396         &  \phantom{$+$}0.00406      \\
    3   &   0.00312         &  \phantom{$+$}0.02289         &  \phantom{$+$}0.01752         &   \phantom{$+$}0.13630         &  \phantom{$+$}0.17808      \\
    4   &   0.00530         &  $-$0.03679                   & $-$0.03811                    &   $-$0.14406                   &  $-$0.13907                \\
    5   &   0.00262         &  $-$0.00497                   & $-$0.00387                    &   $-$0.54697                   &  $-$0.67700                \\
\bottomrule
\end{tabular}
\caption{\textbf{Calculated values of $z_{\ell\lambda}$ and $b_{\ell\lambda}$ for $\beta$\,Cur |} These terms are scaling factors for the polarimetric and photometric amplitudes that take account of the properties of the stellar atmosphere. The values were calculated using the {\tt SYNSPEC/VLIDORT} stellar polarization code \cite{Cotton17, Bailey19, Bailey20b} for a star of $T_{\rm eff} = 27000$\,K and $\log{g} = 3.6$ for wavelengths appropriate to HIPPI-2 (460\,nm), \textit{WIRE} (600\,nm) and \textit{TESS} (800\,nm) for the given $\ell$ values. \\}
\label{tab:zl_bl}
\end{center}
\end{table}

\clearpage
\begin{table}
\begin{center}
\tabcolsep 2. pt
\begin{tabular}{ccccccccccccccccccccccccccccccccc}
\toprule
$\chi^2$ & $i$ & $v\sin i$ & 
$\ell_1$ & $m_1$ & $A_1$ & 
$\ell_2$ & $m_2$ & $A_2$ & 
$\ell_3$ & $m_3$ & $A_3$ & 
$\ell_4$ & $m_4$ & $A_4$ & 
$\ell_5$ & $m_5$ & $A_5$ & 
$\ell_6$ & $m_6$ & $A_6$ & 
$\ell_7$ & $m_7$ & $A_7$ & 
$\ell_8$ & $m_8$ & $A_8$ &  SID  \\
\midrule
 233.5&  42.8  &14&1&$-$1&  9.6   &3&$-$3&   17.5   &4&\phantom{+}0&   15.9&1&1&  6.6   &4&0&    3.5   &4&\phantom{+}0&   17.5   &2&0&    9.7   &4&$-$1&   14.4\\
 233.6&  43.9  &14&1&$-$1& 11.9   &3&$-$3&   14.4   &4&\phantom{+}0&   17.5&1&1&  5.1   &4&0&    8.2   &4&\phantom{+}0&   17.5   &3&$-$1&   17.5   &4&$-$1&   14.4\\
 233.8&  33.9  &14&1&$-$1& 11.9   &4&$-$4&   17.5   &4&\phantom{+}1&   15.9&1&1&  6.6   &4&0&   11.3   &4&\phantom{+}0&   15.9   &2&0&   14.4   &4&$-$1&   15.9\\
 234.2&  37.2  &14&1&$-$1& 11.9   &4&$-$4&   15.9   &4&\phantom{+}0&   15.9&1&1&  5.1   &4&0&    8.2   &4&\phantom{+}0&   17.5   &2&0&   17.5   &4&$-$1&   14.4\\
 234.3&  33.9  &14&1&$-$1& 11.9   &4&$-$4&   14.4   &4&\phantom{+}1&   15.9&1&1&  6.6   &4&0&   11.3   &4&\phantom{+}0&   17.5   &2&0&   14.4   &4&$-$1&   15.9\\
 234.7&  42.8  &14&1&$-$1& 11.9   &3&$-$3&   14.4   &4&\phantom{+}1&   15.9&2&1&  5.1   &4&0&    9.7   &3&$-$3&   11.3   &3&$-$1&    9.7   &4&\phantom{+}0&   14.4   &   B\\
 234.7&  42.8  &14&1&$-$1&  9.6   &3&$-$3&   17.5   &4&\phantom{+}0&   15.9&2&1&  6.6   &4&0&   11.3   &4&\phantom{+}0&   11.3   &3&$-$2&   17.5   &4&$-$1&   14.4\\
 235.2&  47.2  &14&1&$-$1& 10.4   &3&$-$3&   14.4   &4&\phantom{+}0&   23.7&2&1&  5.1   &4&0&    8.2   &3&$-$3&   11.3   &3&$-$3&    9.7   &4&$-$1&   14.4\\
 235.5&  42.8  &14&1&$-$1& 15.7   &3&$-$3&   25.2   &4&\phantom{+}0&   17.5&1&1&  5.1   &4&0&    6.6   &4&\phantom{+}0&   17.5   &3&$-$2&   19.0   &4&$-$1&   19.0\\
 235.6&  42.8  &14&1&$-$1&  9.6   &3&$-$3&   17.5   &4&\phantom{+}0&   15.9&2&1&  5.1   &4&0&   11.3   &4&\phantom{+}0&   15.9   &3&$-$3&    9.7   &4&$-$1&   15.9\\
 235.7&  36.1  &14&1&$-$1& 11.9   &4&$-$4&   14.4   &4&\phantom{+}0&   22.1&1&1&  6.6   &4&0&   11.3   &4&\phantom{+}0&   15.9   &2&0&   17.5   &4&$-$1&   14.4\\
 235.9&  40.6  &14&1&$-$1& 11.9   &3&$-$3&   14.4   &4&\phantom{+}0&   15.9&2&1&  8.2   &4&0&   12.8   &3&$-$3&   15.9   &3&$-$3&    9.7   &4&\phantom{+}0&   15.9\\
 236.0&  36.1  &14&1&$-$1& 11.9   &4&$-$4&   14.4   &4&\phantom{+}0&   15.9&1&1&  6.6   &4&0&   11.3   &4&\phantom{+}0&   14.4   &2&0&   17.5   &4&$-$1&   14.4\\
 236.0&  42.8  &14&1&$-$1&  9.6   &3&$-$3&   17.5   &4&\phantom{+}1&   15.9&2&1&  8.2   &4&0&   11.3   &3&$-$3&   15.9   &3&$-$1&   17.5   &4&$-$1&   15.9   &   C\\
 236.3&  42.8  &14&1&$-$1&  9.6   &3&$-$3&   17.5   &4&\phantom{+}1&   15.9&2&1&  5.1   &4&0&   11.3   &4&\phantom{+}0&   11.3   &3&$-$1&    9.7   &4&$-$1&   15.9\\
 236.3&  42.8  &14&1&$-$1&  9.6   &3&$-$3&   19.0   &4&\phantom{+}1&   15.9&2&1&  9.7   &4&0&   11.3   &3&$-$3&   15.9   &3&$-$1&   17.5   &4&$-$1&   15.9   &   D\\
 236.4&  42.8  &14&1&$-$1&  9.6   &3&$-$3&   17.5   &4&$-$4&   15.9&1&1&  5.1   &4&0&   11.3   &4&\phantom{+}0&   15.9   &3&$-$3&   17.5   &4&$-$1&   15.9\\
 237.4&  40.6  &14&1&$-$1& 11.9   &3&$-$3&   14.4   &4&\phantom{+}0&   15.9&2&1&  6.6   &4&0&   11.3   &3&$-$3&   15.9   &3&$-$3&    9.7   &4&\phantom{+}0&   14.4\\
 237.6&  43.9  &14&1&$-$1&  9.6   &3&$-$3&   15.9   &4&\phantom{+}0&   15.9&2&1&  5.1   &4&0&    9.7   &4&$-$4&   15.9   &3&$-$1&    9.7   &4&$-$1&   15.9\\
 237.7&  42.8  &16&1&$-$1&  9.6   &3&$-$3&   15.9   &4&\phantom{+}0&   15.9&2&1&  5.1   &4&0&   11.3   &4&\phantom{+}0&   11.3   &3&$-$1&    9.7   &4&$-$1&   15.9\\
\vdots &\vdots &\vdots &\vdots &\vdots &\vdots &\vdots &\vdots &\vdots &\vdots &\vdots &\vdots &\vdots& \vdots &\vdots &\vdots &\vdots&\vdots &\vdots &\vdots &\vdots &\vdots &\vdots &\vdots &\vdots &\vdots&\vdots \\
242.6&	55.0&	14&	1&$-$1&	3.5&	3&$-$3&	15.9&	4&\phantom{+}1&	23.7&	2&1&	3.5&	4&0&	11.3&	3&$-$3&	11.3&	3&$-$1&	9.7&	4&\phantom{+}0&	25.2    &   F\\
\vdots &\vdots &\vdots &\vdots &\vdots &\vdots &\vdots &\vdots &\vdots &\vdots &\vdots &\vdots &\vdots& \vdots &\vdots &\vdots &\vdots&\vdots &\vdots &\vdots &\vdots &\vdots &\vdots &\vdots &\vdots &\vdots&\vdots \\
243.1&	55.0&	14&	1&	$-$1&	5.8&	3&	$-$3&	25.2&	4&	\phantom{+}1&	15.9&	2&	1&	3.5&	4&	0&	11.3&	3&	$-$3&	11.3&	3&	$-$1&	17.5&	4&	\phantom{+}0&	25.2    &   G\\
\vdots &\vdots &\vdots &\vdots &\vdots &\vdots &\vdots &\vdots &\vdots &\vdots &\vdots &\vdots &\vdots& \vdots &\vdots &\vdots &\vdots&\vdots &\vdots &\vdots &\vdots &\vdots &\vdots &\vdots &\vdots &\vdots&\vdots \\
246.5&	55.0&	14&	1&$-$1&	11.1&	3&	$-$3&	14.4&	4&	\phantom{+}1&	36.1&	4&	4&	5.1&	4&	0&	5.1&	3&	$-$1&	15.9&	3&	$-$1&	9.7&	4&	\phantom{+}0&	\phantom{0}9.7  &   E\\
\vdots &\vdots &\vdots &\vdots &\vdots &\vdots &\vdots &\vdots &\vdots &\vdots &\vdots &\vdots &\vdots& \vdots &\vdots &\vdots &\vdots&\vdots &\vdots &\vdots &\vdots &\vdots &\vdots &\vdots &\vdots &\vdots&\vdots \\
246.7&	55.0&	16&	1&	$-$1&	11.1&	3&	$-$3&	14.4&	4&	\phantom{+}1&	36.1&	4&	4&	5.1&	4&	0&	3.5&	3&	$-$3&	15.9&	3&	$-$1&	9.7&	4&	\phantom{+}0&	9.7 &   H\\
\vdots &\vdots &\vdots &\vdots &\vdots &\vdots &\vdots &\vdots &\vdots &\vdots &\vdots &\vdots &\vdots& \vdots &\vdots &\vdots &\vdots&\vdots &\vdots &\vdots &\vdots &\vdots &\vdots &\vdots &\vdots &\vdots&\vdots \\
373.3&	42.8&	18&	1&	\phantom{+}0&	12.7&	3&	$-$3&	28.3&	4&	\phantom{+}1&	17.5&	4&	4&	36.1&	4&	0&	22.1&	3&	$-$3&	40.7&	3&	\phantom{+}0&	42.3&	3&	$-$1&	23.7    &   I\\
\vdots &\vdots &\vdots &\vdots &\vdots &\vdots &\vdots &\vdots &\vdots &\vdots &\vdots &\vdots &\vdots& \vdots &\vdots &\vdots &\vdots&\vdots &\vdots &\vdots &\vdots &\vdots &\vdots &\vdots &\vdots &\vdots&\vdots \\
407.9&	46.1&	16&	1&	$-$1&	22.6&	3&	$-$3&	23.7&	2&	\phantom{+}0&	15.9&	2&	1&	33.0&	2&	0&	3.5&	3&	$-$3&	48.5&	4&	\phantom{+}0&	19.0&	3&	$-$1&	42.3    &   A\\
\vdots &\vdots &\vdots &\vdots &\vdots &\vdots &\vdots &\vdots &\vdots &\vdots &\vdots &\vdots &\vdots& \vdots &\vdots &\vdots &\vdots&\vdots &\vdots &\vdots &\vdots &\vdots &\vdots &\vdots &\vdots &\vdots&\vdots \\
\bottomrule
\end{tabular}
\caption{\textbf{Best common solutions for the identification of
    the modes with frequencies $f_1, \ldots, f_8$\ |} Listed are the
  20 best parameter combinations, along with those that survive later culling from additional polarimetric constraints (identified by SID designations, defined later), among thousands resulting from the mode
  identification for the eight modes from the Fourier Parameter Fit method. We have used a parameter grid-based approach coupled to a genetic algorithm applied to the measured amplitude and
  phase distributions across the Si\,III 4552.6\,\AA\ line profile of
  $\beta\,$Cru shown in Figure\,\ref{fig:lpa}. The inclination angle is expressed in degrees; $v\sin i$ and the mode amplitudes are given in km\,s$^{-1}$.
  The parameters were allowed to
  vary as $\ell_j\in [0,4]$ and $m_j \in [-\ell_j,\ell_j]$ for $j=1, \ldots, 8$,
while the mode amplitudes, $A_j\in [1,50]\,$km\,s$^{-1}$ with a step
  of 1\,km\,s$^{-1}$. From all of the line profile fits based on each of the
  individual eight modes, this table was extracted by demanding a single value
  for the inclination angle
$i\in [0^\circ,75^\circ]$ in stpdf of $2^\circ$,
  the projected surface rotation velocity $v\sin i\in [10,40]$\,km\,s$^{-1}$ in
  stpdf of 2\,km\,s$^{-1}$ and the intrinsic line profile in the absence of
  pulsations and rotation (approximated by a Gaussian shape with width
  $\sigma_{\rm intr}\in [10,30]$\,km\,s$^{-1}$ in stpdf of
  2\,km\,s$^{-1}$). 
  Note that the mode amplitudes given here in this table
are ``local'' maximal velocity values due to the local eigenvector displacements
of the mode, while those in Table \ref{tab:frequencies} are the net effect
obtained after integration over the visible limb-darkened hemisphere in the
line of sight.
For all the listed solutions, $\sigma_{\rm intr}=26\,$km\,s$^{-1}$ was found. We find that the identification of the dominant mode in the spectroscopy and polarimetry, $(l_2,m_2)=(3,-3)$ or $(4,-4)$ is almost equivalent. This can be understood in terms of the slow rotation and the inclination angle of the star.
We show the distribution of $i$ and $v\sin i$ for all the solutions having
$\chi^2\leq\,1000$ in Figure\,\ref{fig:inclvsini}. In this figure, we 
also include insets for the distribution of solutions compliant with the inclination restrictions for each mode imposed by the polarimetry. The solutions represented in the two insets deliver  
$i=43^\circ\pm3^\circ$ and $v\sin i=14\pm2\,$km\,s$^{-1}$ for the overall best solution having $(l_2,m_2)=(3,-3)$.
\\}
\label{tab:MI-all8}
\end{center}
\end{table}

\clearpage
\begin{table}
\begin{center}
\tabcolsep 3 pt
\resizebox{\textwidth}{!}{%
\renewcommand{\arraystretch}{1.25}
\begin{tabular}{cccccccccccccccc}
\toprule
ID  &   &   $f_1$   &   $f_2$   &   $f_3$   &   $f_4$   &   $f_5$   &   $f_6$   &   $f_7$   &   $f_8$   \\
$f_{\textit{T+W}}$  &   (c\,d$^{-1}$)   &   5.229776(1) &   5.96364(2)\0    &   5.477750(3)\0   &   5.2162(6)\0\0   &   5.6977(3)\0\0    &   7.6605(3)\0\0   &   6.3596(6)\0\0   &   5.9229(6)\0\0   \\
$A_{\textit{TESS}}$ &   (ppm)           &   3639(16)    &   \0231(10)       &   1641(10)        &    \0144(10)      &   \0119(10)         &   \0143(9)\0      &   \0\089(9)\0     &   \0\090(9)\0 \\      
\midrule
SID & $i$ ($^{\circ}$) &    
$(\ell_1,\,m_1)$ &
$(\ell_2,\,m_2)$ &
$(\ell_3,\,m_3)$ &
$(\ell_4,\,m_4)$ &  
$(\ell_5,\,m_5)$ &  
$(\ell_6,\,m_6)$ &  
$(\ell_7,\,m_7)$ &  
$(\ell_8,\,m_8)$ \\  
&   $v\sin i$   &
$A^M_Q$ (ppm)  &
$A^M_Q$ (ppm)  &
$A^M_Q$ (ppm)  &
$A^M_Q$ (ppm)  &
$A^M_Q$ (ppm)  &
$A^M_Q$ (ppm)  &
$A^M_Q$ (ppm)  &
$A^M_Q$ (ppm)  \\
&   ($km\,s^{-1}$) &
$A^M_U$ (ppm)  &
$A^M_U$ (ppm)  &
$A^M_U$ (ppm)  &
$A^M_U$ (ppm)  &
$A^M_U$ (ppm)  &
$A^M_U$ (ppm)  &
$A^M_U$ (ppm)  &
$A^M_U$ (ppm)  \\
\midrule
A  &	46.1    &	$(1,-1)$    &   $(3,-3)$    &   $(2,\phantom{+}0)$    &   $(2,+1)$    &   $(2,\phantom{+}0)$  &   $(3,-3)$    &   $(4,\phantom{+}0)$    &   $(3,-1)$ \\
        & 16          &   $0.\phantom{00\substack{+0.00\\+0.00}}$  &   \textbf{$5.87\substack{+0.62\\-0.56}$}  &   $1.95\substack{+0.48\\-0.78}$           &   $0.05\substack{+0.00\\-0.00}$   &   $0.14\substack{+0.04\\-0.06}$   &   $3.63\substack{+0.44\\-0.40}$   &   $2.64\substack{+0.27\\-0.27}$   &   $0.42\substack{+0.06\\-0.07}$   \\
        &           &   $0.\phantom{00\substack{+0.00\\+0.00}}$  &   \textbf{$5.49\substack{+0.65\\-0.59}$}  &   $0.\phantom{00\substack{+0.00\\+0.00}}$ &   $0.07\substack{+0.01\\-0.01}$   &   $0.\phantom{00\substack{+0.00\\-0.00}}$   &   \textbf{$3.40\substack{+0.44\\-0.41}$}   &   $0.\phantom{00\substack{+0.00\\-0.00}}$   &   $1.32\substack{+0.17\\-0.19}$   \\
\hdashline[1pt/5pt]
 B  &   42.8    &   $(1,-1)$    &   $(3,-3)$    &   $(4,+1)$    &   $(2,+1)$    &   $(4,\phantom{+}0)$  &   $(3,-3)$    &   $(3,-1)$    &   $(4,\phantom{+}0)$ \\
        & 14          &   $0.\phantom{00\substack{+0.00\\+0.00}}$  &   \textbf{$6.85\substack{+0.78\\-0.69}$}    &   $0.26\substack{+0.24\\-0.25}$ &    $0.05\substack{+0.00\\-0.00}$   &   $3.67\substack{+0.38\\-0.33}$   &   $4.24\substack{+0.54\\-0.49}$    &   $0.48\substack{+0.06\\-0.06}$    &   $2.77\substack{+0.33\\-0.29}$  \\     
        &           &   $0.\phantom{00\substack{+0.00\\+0.00}}$  &   \textbf{$6.54\substack{+0.81\\-0.72}$}    &   $2.09\substack{+0.21\\-0.20}$ &    $0.07\substack{+0.01\\-0.01}$   &   $0.\phantom{00\substack{+0.00\\+0.00}}$ &   \textbf{$4.05\substack{+0.55\\-0.50}$}    &   $1.15\substack{+0.14\\-0.15}$    &   $0.\phantom{00\substack{+0.00\\+0.00}}$    \\
\hdashline[1pt/5pt]
 C  &   42.8    &   $(1,-1)$    &   $(3,-3)$    &   $(4,+1)$    &   $(2,+1)$    &   $(4,\phantom{+}0)$  &   $(3,-3)$    &   $(3,-1)$    &   $(4,-1)$    \\
        & 14          &   $0.\phantom{00\substack{+0.00\\+0.00}}$  &   \textbf{$6.85\substack{+0.78\\-0.69}$}    &   $0.26\substack{+0.24\\-0.25}$ &    $0.05\substack{+0.00\\-0.00}$   &   $3.67\substack{+0.38\\-0.33}$   &   $4.24\substack{+0.54\\-0.49}$    &   $0.48\substack{+0.06\\-0.06}$    &   $0.02\substack{+0.02\\-0.02}$  \\     
        &           &   $0.\phantom{00\substack{+0.00\\+0.00}}$  &   \textbf{$6.54\substack{+0.81\\-0.72}$}    &   $2.09\substack{+0.21\\-0.20}$ &    $0.07\substack{+0.01\\-0.01}$   &   $0.\phantom{00\substack{+0.00\\+0.00}}$ &   \textbf{$4.05\substack{+0.55\\-0.50}$}    &   $1.15\substack{+0.14\\-0.15}$    &   $0.18\substack{+0.02\\-0.02}$    \\
\hdashline[1pt/5pt]
 D  &   42.8    &   $(1,-1)$    &   $(3,-3)$    &   $(4,+1)$    &   $(2,+1)$    &   $(4,\phantom{+}0)$  &   $(3,-3)$    &   $(3,-1)$    &   $(4,-1)$    \\
        & 14          &   $0.\phantom{00\substack{+0.00\\+0.00}}$  &   \textbf{$6.85\substack{+0.78\\-0.69}$}    &   $0.26\substack{+0.24\\-0.25}$ &    $0.05\substack{+0.00\\-0.00}$   &   $3.67\substack{+0.38\\-0.33}$   &   $4.24\substack{+0.54\\-0.49}$    &   $0.48\substack{+0.06\\-0.06}$    &   $0.02\substack{+0.02\\-0.02}$  \\     
        &           &   $0.\phantom{00\substack{+0.00\\+0.00}}$  &   \textbf{$6.54\substack{+0.81\\-0.72}$}    &   $2.09\substack{+0.21\\-0.20}$ &    $0.07\substack{+0.01\\-0.01}$   &   $0.\phantom{00\substack{+0.00\\+0.00}}$ &   \textbf{$4.05\substack{+0.55\\-0.50}$}    &   $1.15\substack{+0.14\\-0.15}$    &   $0.18\substack{+0.02\\-0.02}$    \\
\hdashline[1pt/5pt]
 E  &	55.0    &	$(1,-1)$    &   $(3,-3)$    &   $(4,+1)$    &   $(2,+1)$    &   $(4,\phantom{+}0)$  &   $(3,-1)$    &   $(3,-1)$    &   $(4,-1)$    \\
        & 14          &   $0.\phantom{00\substack{+0.00\\+0.00}}$  &  \textbf{$4.07\substack{+0.36\\-0.33}$}   &   $1.41\substack{+0.13\\-0.10}$   &   $0.05\substack{+0.00\\-0.00}$   &   $4.22\substack{+0.47\\-0.56}$   &   $0.04\substack{+0.18\\-0.44}$   &   $0.03\substack{+0.12\\-0.28}$   &   $0.08\substack{+0.01\\-0.01}$   \\
        &           &   $0.\phantom{00\substack{+0.00\\+0.00}}$  &  \textbf{$3.52\substack{+0.40\\-0.36}$}   &   $0.94\substack{+0.17\\-0.17}$   &   $0.09\substack{+0.01\\-0.01}$   &   $0\phantom{.00\substack{+0.00\\-0.00}}$ &   \textbf{$3.77\substack{+0.68\\-1.04}$}   &   $2.40\substack{+0.46\\-0.69}$   &   $0.05\substack{+0.01\\-0.01}$   \\
\hdashline[1pt/5pt]
 F  &	55.0    &   $(1,-1)$    &   $(3,-3)$    &   $(4,+1)$    &   $(2,+1)$    &   $(4,\phantom{+}0)$  &   $(3,-3)$    &   $(3,-1)$    &   $(4,-1)$    \\
        & 14          &   $0.\phantom{00\substack{+0.00\\+0.00}}$  &  \textbf{$4.07\substack{+0.36\\-0.33}$}   &   $1.41\substack{+0.13\\-0.10}$   &   $0.05\substack{+0.00\\-0.00}$   &   $4.22\substack{+0.47\\-0.56}$   &   $2.52\substack{+0.26\\-0.25}$   &   $0.03\substack{+0.11\\-0.27}$   &   $0.08\substack{+0.01\\-0.01}$   \\
        &           &   $0.\phantom{00\substack{+0.00\\+0.00}}$  &  \textbf{$3.52\substack{+0.40\\-0.36}$}   &   $0.94\substack{+0.17\\-0.17}$   &   $0.09\substack{+0.01\\-0.01}$   &   $0.\phantom{00\substack{+0.00\\-0.00}}$ &   \textbf{$2.18\substack{+0.27\\-0.26}$}   &   $2.35\substack{+0.46\\-0.67}$   &   $0.05\substack{+0.01\\-0.01}$   \\
\hdashline[1pt/5pt]
 G  &	55.0    &	$(1,-1)$    &   $(3,-3)$    &   $(4,+1)$    &   $(2,+1)$    &   $(4,\phantom{+}0)$  &   $(3,-3)$    &   $(3,-1)$    &   $(4,-1)$    \\
        & 14          &   $0.\phantom{00\substack{+0.00\\+0.00}}$  &  \textbf{$4.07\substack{+0.36\\-0.33}$}   &   $1.41\substack{+0.13\\-0.10}$   &   $0.05\substack{+0.00\\-0.00}$   &   $4.22\substack{+0.47\\-0.56}$   &   $2.52\substack{+0.26\\-0.25}$   &   $0.03\substack{+0.11\\-0.27}$   &   $0.08\substack{+0.01\\-0.01}$   \\
        &           &   $0.\phantom{00\substack{+0.00\\+0.00}}$  &  \textbf{$3.52\substack{+0.40\\-0.36}$}   &   $0.94\substack{+0.17\\-0.17}$   &   $0.09\substack{+0.01\\-0.01}$   &   $0.\phantom{00\substack{+0.00\\-0.00}}$ &   \textbf{$2.18\substack{+0.27\\-0.26}$}   &   $2.35\substack{+0.46\\-0.67}$   &   $0.05\substack{+0.01\\-0.01}$   \\
\hdashline[1pt/5pt]
 H  &	55.0    &	$(1,-1)$    &   $(3,-3)$    &   $(4,+1)$    &   $(2,+1)$    &   $(4,\phantom{+}0)$  &   $(3,-3)$    &   $(3,-1)$    &   $(4,-1)$    \\
        & 16          &   $0.\phantom{00\substack{+0.00\\+0.00}}$  &  \textbf{$4.07\substack{+0.36\\-0.33}$}   &   $1.41\substack{+0.13\\-0.10}$   &   $0.05\substack{+0.00\\-0.00}$   &   $4.22\substack{+0.47\\-0.56}$   &   $2.52\substack{+0.26\\-0.25}$   &   $0.03\substack{+0.11\\-0.27}$   &   $0.08\substack{+0.01\\-0.01}$   \\
        &           &   $0.\phantom{00\substack{+0.00\\+0.00}}$  &  \textbf{$3.52\substack{+0.40\\-0.36}$}   &   $0.94\substack{+0.17\\-0.17}$   &   $0.09\substack{+0.01\\-0.01}$   &   $0.\phantom{00\substack{+0.00\\-0.00}}$ &   \textbf{$2.18\substack{+0.27\\-0.26}$}   &   $2.35\substack{+0.46\\-0.67}$   &   $0.05\substack{+0.01\\-0.01}$   \\
\midrule
 I  &	42.8    &	$(1,\phantom{+}0)$    &   $(3,-3)$    &   $(4,+1)$    &   $(4,+4)$    &   $(4,\phantom{+}0)$  &   $(3,-3)$    &   $(3,\phantom{+}0)$    &   $(3,-1)$ \\
        & 18          &   $0.\phantom{00\substack{+0.00\\+0.00}}$  &   \textbf{$6.85\substack{+0.78\\-0.69}$}    &   $0.26\substack{+0.24\\-0.25}$ &    $2.22\substack{+0.28\\-0.26}$   &   $3.67\substack{+0.38\\-0.33}$   &   $4.24\substack{+0.54\\-0.49}$    &   $5.93\substack{+6.64\\-1.95}$    &   $0.49\substack{+0.06\\-0.06}$  \\     
        &           &   $0.\phantom{00\substack{+0.00\\+0.00}}$  &   \textbf{$6.54\substack{+0.81\\-0.72}$}    &   $2.09\substack{+0.21\\-0.20}$ &    $2.12\substack{+0.29\\-0.26}$   &   $0.\phantom{00\substack{+0.00\\+0.00}}$ &   \textbf{$4.05\substack{+0.55\\-0.50}$}    &   $0.\phantom{00\substack{+0.00\\+0.00}}$    &   $1.16\substack{+0.14\\-0.15}$    \\
\bottomrule
\end{tabular}}
\renewcommand{\arraystretch}{1.0}
\caption{\textbf{Theoretical polarimetric amplitudes of best solutions |} For each frequency with a spectroscopic detection we use the \textit{TESS} amplitude and the given inclination to calculate the expected polarimetric amplitudes for the given modes from the analytical model \cite{Watson83} (we estimate 1-$\sigma$ errors based on the errors in $A_{TESS}$ and $\pm$\,2$^\circ$ in inclination). Note that these determinations are strictly valid only for the time period covered by the \textit{TESS} observations. From the solutions given in Supplementary Table \ref{tab:MI-all8} modes with predicted polarimetric amplitudes $>5.5$\,ppm and non-detections are eliminated, as are those detected, but with predicted polarimetric amplitudes $<2.0$\,ppm; the remaining solutions are listed in order of plausibility based on the observed and predicted polarimetric amplitudes and the $\chi^2$ from spectroscopic mode identification. The table shows eight solutions that strictly meet these criteria, and another one (below the solid line) where the 1-$\sigma$ errors and/or rounding the predicted amplitudes to 1 decimal place meets the criteria.} 
\label{tab:MI-best}
\end{center}
\end{table}


\end{document}